\documentclass[amsmath,amssym,amsfonts,pre,twocolumn,showpacs]{revtex4}
\usepackage[dvips]{epsfig}

\newcommand{\rv}{{\bf r}}
\newcommand{\rvh}{\hat{{\bf r}}}
\newcommand{\nv}{\hat{{\bf n}}}
\newcommand{\bv}{{\bf b}}

\newcommand{\xh}{\hat{x}}
\newcommand{\yh}{\hat{y}}
\newcommand{\zh}{\hat{z}}
\newcommand{\xv}{{\bf x}}

\newcommand{\uv}{{\bf u}}

\newcommand{\tv}{\hat{{\bf t}}}

\newcommand{\bD}{{\bf \Delta}}
\newcommand{\grad}{{\bf \nabla}}

\begin{document}

\title{Defects in Crystalline Packings of Twisted Filament Bundles:  I. Continuum Theory of Disclinations}
\author{Gregory M. Grason}
\affiliation{Department of Polymer Science and Engineering, University of Massachusetts, Amherst, MA 01003, USA}
\begin{abstract}

We develop the theory of the coupling between in-plane order and out-of-plane geometry in twisted, two-dimensionally ordered filament bundles based on the non-linear continuum elasticity theory of columnar materials.  We show that twisted textures of filament backbones necessarily introduce stresses into the cross-sectional packing of bundles and that these stresses are formally equivalent to the geometrically-induced stresses generated in thin elastic sheets that are forced to adopt spherical curvature.  As in the case of crystalline order on curved membranes, geometrically-induced stresses couple elastically to the presence of topological defects in the in-plane order.  We derive the effective theory of multiple disclination defects in the cross section of bundle with a fixed twist and show that above a critical degree of twist, one or more 5-fold disclinations is favored in the elastic energy ground state.  We study the structure and energetics of multi-disclination packings based on models of equilibrium and non-equilibrium cross-sectional order.
\end{abstract}
\date{\today}

\maketitle
\section{Introduction}

Topological defects are crucial components of the structure and thermodynamics in many frustrated systems in condensed matter~\cite{nelson_defects}.  Perhaps the most well-known example of frustration-induced defects is exemplified by the Abrikosov phase of type-II superconductors where the application of a external magnetic field leads to the proliferation of vortices where the superconducting order is destroyed and through which magnetic flux is threaded~\cite{abrikosov}.  One remarkable feature of such frustrated systems is appearance of defects in the ground state, indicating that the balance of competing thermodynamic forces requires highly heterogeneous distributions of energy and structure.  In {\it geometrically frustrated} systems, frustration arises intrinsically from the incompatibility of locally-preferred order with geometrical constraints on long-range ordering~\cite{sadoc_frustration}.  Examples of geometrical frustration abound in self-organized molecular and colloidal systems.  In bulk, three-dimensional materials, well known examples include liquid-crystal blue phases~\cite{wright_mermin} as well as the twist-grain boundary phases of chiral smectics~\cite{renn_lubensky}.   When confined to two-dimensional surfaces, materials possessing a range of anisotropic order---nematic~\cite{lubensky_prost, lubensky_mackintosh, vitelli_turner}, hexatic~\cite{nelson_peliti}, smectic~\cite{santangelo, santangelo_pre}, crystalline~\cite{nelson_peliti, nelson_seung, bowick_travesset_nelson, bowick_caccuito, bausch_science, irvine_nature}---are generically frustrated by the presence of non-zero Gaussian curvature, which couples favorably to the presence of disclinations in the ground states.  

Condensed phases of chiral polymers are subject to a unique frustration between two-dimensional (2D) ordering perpendicular to the chain backbone and a microscopic tendency for intermolecular twist.  Kamien and Nelson showed that in bulk columnar materials chirality gives rise to two distinct types of grain-boundary phases---the tilt-grain boundary and moir\'e phases---both composed of arrays of dislocations lying in the plane of 2D order~\cite{kamien_nelson_95, kamien_nelson_96}.  Underlying the frustration of bulk columnar phases of chiral polymers is the geometrical coupling between in-plane displacements and filament tilts:  inter-filament twist requires gradients in in-plane shear stresses along the molecular backbone.  The frustration of chiral polymer assemblies is particularly important to the assembly of biological polymers, from extra-cellular proteins like collagen~\cite{wess} and fibrin~\cite{weisel} to the cytoskeletal filaments, f-actin~\cite{bausch_pnas_08, shin_grason_prl_09}.  These are universally helical molecules and are organized into densely-packed states in cells and tissues  of living organisms~\cite{bouligand}.  A number of studies show that the frustration between 2D and chiral ordering gives rise to important thermodynamic properties including self-limiting assembly of bundles and fibers of chiral filaments~\cite{weisel_pnas_87, turner_prl_03, grason_bruinsma_07, grason_09, heussinger, hagan_prl_10}.  

Recently, we have demonstrated that frustration between chiral order and 2D packing leads to another important and surprising consequence, the restructuring of the ground state packing of twisted filament bundles~\cite{grason_prl_10}.  In particular, we reported that twisted bundles are akin to 2D crystals formed on the spherically-curved surfaces, in which out-of-plane geometry generates in-plane stresses that are partially screened in the ground state by topological defects, 5-fold disclinations, in the lattice order of bundle cross sections.  Hence, it was proven that ground state in-plane order of sufficiently twisted bundles is irregular, possessing an excess of topological defects in the configurations that minimize the elastic energy of filament packing.

\begin{figure}
\center \epsfig{file=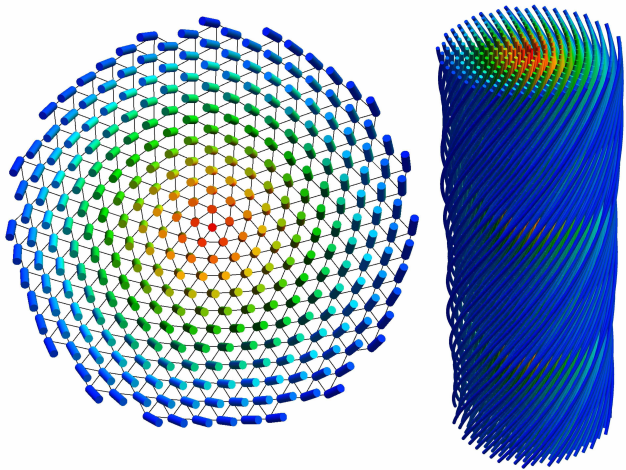, width=3.35in}\caption{ The top (left) and side (right) view of a twisted filament bundle possessing and 5-fold disclination in the center.  The filaments are colored to denote the tilt of filaments with respect to the center axes, with red filaments parallel the rotation axis and blue filaments most inclined. Here, the reduced twist corresponds to $\Omega R \simeq 0.7$}
\label{fig: side_top}
\end{figure}

In this article we describe in detail the theory of disclinations in twisted filament bundles and its derivation from non-linear continuum elasticity theory of 2D-ordered filament arrays.  The unique symmetries exhibited by these materials lead to an intrinsic, non-linear coupling between in-plane strains and out-of-plane deflections of the filament backbone:  inter-filament twist makes it impossible to evenly space filaments in cross section.  This analysis of the non-linear theory shows that the equations of mechanical equilibrium in bundles are subject a compatibility condition linking twist of filament backbones and in-plane stresses.  In analogy to the F\"oppl-von K\'arman theory of elastic plates and membranes, this compatibility condition for filament bundles shows that twist generates in-plane stresses in bundles equivalent to those generated in a positively-curved elastic sheet possessing a Gaussian curvature equal to $3 \Omega^2$, where $2 \pi /\Omega$ is the helical pitch of the bundle.  Motivated by the fact that topological defects screen curvature-induced stresses in crystalline membranes, we derive the effective theory for disclinations in the cross section of twisted bundles.  We show that above a critical of twist $(\Omega R)_c = \sqrt{2/9}$, 5-fold disclinations are trapped in the cross section, where $R$ is the bundle radius.  An example of a twisted bundle ground state possessing an energetically-favorable 5-fold disclination is shown in Fig.~\ref{fig: side_top}.  At higher values of twist, multiple disclinations are stabilized in the cross section of crystalline bundle packings.  Based on the continuum theory, we study the discrete spectrum of elastic-energy minimizing configurations as a function of bundle twist which efficiently screen the buildup of twist-induced stresses.  Further, we show that for a simple model surface-nucleated defects,  the relatively weak repulsions between disclinations near the free boundaries of twisted bundles lead to a ``pile up" of 5-fold defects    just above the critical size for disclination stability.  The excess defects entrapped within these non-equilibrium structures suggest that a limited mobility of disclinations within the bulk of bundles drastically reduces the screening of geometrically-induced stresses in ``surface-grown" bundles.  

The remainder of this article is organized as follows.  In Sec. \ref{sec: strain} describe the rotationally-invariant properties of the non-linear elasticity theory of 2D ordered filament arrays (i.e. columnar order).   In Sec. \ref{sec: mecheq} we derive and study the equations of mechanical equilibrium for twisted bundles in the presence disclinations in the cross section, showing that one or more disclinations become stabilized by twisted-induced stress above a critical twist.  In Sec. \ref{sec: multi} we present the results for multi-disclination cross sections in twisted bundles for both ground-state packings and a model of non-equilibrium bundle assembly.  We conclude with a brief discussion of the surprising correspondence between crystalline order on spherically-curved surfaces and crystalline order of twisted bundles.  Appendix A describes the coarse-graining of inter-filament interactions and the form of the non-linear elastic energy of multi-filament assemblies.  An important result of this study is the exact and closed-form solution for the elastic energy of multiple disclinations in a 2D crystal, bounded by a free cylindrical surface.  To our knowledge, no such solution has been previously published.  As these results are more broadly applicable to finite-sized crystalline materials beyond the specific context of twisted filament bundles we include the full details of the derivation of the disclination-induced stresses and energies in Appendix B.  In a subsequent paper~\cite{azadi_grason}, we analyze the role of dislocations---``neutral", 5-7 pairs of disclinations---in ground-state order of twisted bundles as well as the structure and thermodynamics of multi-dislocation groundstates of twisted bundles.

\section{Rotational Invariance and the Non-Linear Strain of Ordered Filament Arrays}
\label{sec: strain}

In this section we construct the elastic energy that describes deformations of an ordered array of filamentous elements. We assume the stress-free reference state of this energy to be an equidistant hexagonal array of uniformly straight filaments, aligned along the $\zh$ direction.  To describe deformations, we consider the mapping of a material point filament array in its reference configuration, $\xv = \xv_\perp + z_0 \hat{z}$, to the position of the same point in the deformed configuration, $\rv(\xv)= \rv_\perp(\xv)+z \hat{z}$, where $\xv_\perp$ denotes the in-plane components $\xv$.  We the define the in-plane displacement $\uv(\xv) = \rv_\perp(\xv) - \xv_\perp$, which the $xy$ displacement of a point which, after deformation, is in a plane at $z$.  Here, it is most convenient to adopt a ``mixed representation" where we define $u(\xv)$ as function of material coordinates, $\xv_\perp$, in the plane (a Lagrangian variable) and a function ``current" vertical position, $z$ (a Eulerian variable)~\cite{stenull_lubensky}.  

As filamentous assemblies possess a type of two-dimensional order, it is essential to construct the strain in terms of a two-dimensional strain tensor that describes changes in inter-filament spacing in a plane perpendicular to the backbone orientation or chain tangent, $\tv$.  The tangent is related to the displacement by $\tv \simeq \zh +\partial_z \uv_\perp$~\cite{bruinsma_selinger}. This property is necessary to ensure that deformations that slide filaments along their long axis cost no mechanical energy.  The elastic energy of a material with this type of columnar order~\cite{degennes_prost} has the following form
\begin{equation}
\label{eq: elastic}
E = \frac{1}{2} \int dV \big( \lambda u_{kk}^2 + 2 \mu u_{ij} u_{ij} \big),
\end{equation}
where $u_{ij}$ is the elastic strain tensor which has components in the $x$ and $y$ direction only.  Here, $\lambda$ and $\mu$ are the Lam\'e elastic coefficients characterizing the in-plane, isotropic elastic properties of the hexagonal filament lattice.  The precise form of the strain tensor must be such that uniform rigid rotations of the reference state of the filament array require no elastic energy.  The form of the strain satisfying these conditions is,
\begin{equation}
\label{eq: uij}
u_{ij} = \frac{1}{2} \big( \partial_i u_j + \partial_j u_i + \partial_i \uv \cdot \partial_j \uv  - \partial_z u_i \partial_z u_j \big).
\end{equation}
In addition to the standard symmetric derivatives in the linear elastic strain, $u_{ij}$ includes two non-linear contributions which ensure rotational invariance of the strain.  The combination of the first 3 terms in eq. (\ref{eq: uij}), familiar to the non-linear strain of 2D solids, is invariant under rotations around the $\zh$ axis~\cite{landau_lifshitz}.   The the final term is necessary to preserve rotational invariance under rotations around an axis in the $xy$ plane~\cite{bruinsma_selinger}.  It is straightforward to show, for example, that a rigid rotation around the $\yh$ axis by an angle by $\theta$ described by, $u_x= x (1 - \sec \theta)  + z \tan \theta $ leads to vanishing strain.

This final non-linear contribution to the strain in eq. (\ref{eq: uij}) from the out-of-plane derivatives of $\uv$ derives from the {\it layered geometry} of the filament array.  The filament array can be decomposed into a series of rows, ruled surfaces, stacked along directions in the plane perpendicular to the filament tangent.  And not unlike the non-linear strain of fluid membrane stacks (i.e. smectics)~\cite{grinstein_pelcovits}, the elastic energy is insensitive to sliding these rows along the tangent direction, and hence, the non-linear form of the strain ensures that the elastic energy measures distances perpendicular to $\tv$.   That is, if neighbor filaments are separated by $\bD$ in the $xy$ plane, the non-linear strain is sensitive to the perpendicular projection of this distance, or 
\begin{equation}
\bD_\perp = \bD - \tv (\tv \cdot \bD) .
\end{equation}
which measures the {\it distance of closest approach} between the filament pair.   In Appendix A, for microscopic model in which filaments are treated as continuous curves whose segments interact via a central force law (such as a Lennard-Jones potential) we show that the potential energy of the system is most naturally expressed as function of $\bD_\perp$ alone.  Furthermore, by explicitly constructing distortions of this microscopic model from its mechanically-stable, hexagonally-ordered reference state, we find that the course-grained energy is identical to eq. (\ref{eq: elastic}) described by the non-linear elastic strain tensor, $u_{ij}$, of the form eq. (\ref{eq: uij}).  Thus, the contribution from  $-\partial_z u_i \partial_z u_j \simeq -t_i t_j$ to the form of the non-linear strain may be viewed as the result of the projection of inter-filament distance changes in a plane perpendicular to $\tv$, i .e. $\delta |\bD_\perp|^2 = 2 u_{ij} \delta x_i \delta x_j$.

For the case of small strains, it is often sufficient to maintain the linearized strain tensor, $u_{ij} \simeq (\partial_i u_j + \partial_j u_j)/2$.  However, as materials with columnar order are particular ``soft" to tilt or bend deformations due to the lack of out-of-plane shear response, in-plane tilt deformations become significant in many situations.  It has been demonstrated that the anharmonic coupling between in-plane strains and filament tilt gives rise to out-of-plane buckling instabilities in columnar systems~\cite{bruinsma_selinger, kleman_ostwald}.  In particular, when subject to a sufficiently large uniform, uniaxial stress in the cross section of the material, the columns become unstable to certain long-wavelength modes along the long axes of the cylinder, which reduce inter-filament spacing at the expense of out-of-plane bending.  Hence, these filamentous and columnar materials are subject to the same Helfrich-Hurault instability well known in smectic layer systems~\cite{chaikin_lubensky}.  In the following, we show that this instability, is a particular example of a more general phenomena in which stress in the plane of filament order couples to out-of-plane deformations of flexible filament arrays.   Due to the generic form of the non-linear coupling between filament tilt and elastic strain, we find that certain configurations of $\tv$ necessarily introduce stress into the array, in a way that is formally quite similar to the relationship between in-plane stress and the Gaussian curvature of thin elastic sheets.  For the purposes of the present study, we focus on the case of superhelically twisted bundles for which we demonstrate that the elastic energy ground state is, in general, highly complex and riddled with energy-minimizing configurations of topological defects in the lattice order of the bundle cross section. 

While it is important to keep in mind that is necessary to retain the contribution from $\partial_i \uv \cdot \partial_j \uv$ to the strain for large in-plane rotations that necessarily occur between distant planar cross sections of the twisted array, in the following we will drop the contribution from $\partial_i \uv_\perp \cdot \partial_j \uv_\perp$ when referring to $u_{ij}$.  Further, we use the fact that $t_i \simeq \partial_z u_i$ to write the non-linear strain more simply as 
\begin{equation}
\label{eq: ut}
u_{ij} \simeq \frac{1}{2} \big( \partial_i u_j + \partial_j u_i - t_i t_j \big),
\end{equation}
highlighting the geometrical coupling between in-plane strains and out-of-plane geometry unique to the non-linear theory of columnar materials.

\section{Mechanical Equilibrium in Twisted Bundles}

\label{sec: mecheq}

In this section we derive the equations governing the mechanical equilibria of filament bundles that are superhelically twisted.  Our primary task is to demonstrate that a necessary consequence of this apparently simple geometry is a non-uniform distribution of the stress in the cross section of the bundle.  Hence, the appearance of stress in these bundles is an indication of the fact that it is geometrically impossibly to maintain a uniform spacing of filaments in the presence of twist~\cite{starostin}.  Our formulation of the twist-induced stresses in bundles casts this geometric frustration in a surprising light.  From the point of the view of the stress distribution, global twist of the bundle acts like uniform distribution of {\it disclinations} in the cross section whose density increases with the square of the bundle rotation rate.  Thus, our second task in this section is to demonstrate the energetic coupling between the presence of twist-induced stress and topological defects in the lattice order of the filaments.  

\subsection{Equations of mechanical equilibrium}

We consider a cylindrical bundle subject to a uniform rate, $\Omega$, of twist around the central axis of the bundle.  Additionally, we assume that the cross sections of the bundle are identical up to a rotation.  In this case, we decompose the displacements into two steps:  1) $\uv (\xv_\perp)$, the in-plane displacement field from hexagonal order of a given reference plane of the bundle, say $z=0$; and 2) $\uv_\Omega (\xv_\perp, z)$, the overall composite displacement resulting from helical rotation of positions $\xv'_\perp=\xv_\perp + \uv$ at a constant rate, $\Omega$, around $\zh$ along the bundle.
\begin{equation}
\label{eq: rot}
\uv_\Omega(\xv) = \big[ \cos (\Omega z)  - 1\big] (x' \xh + y' \yh) -\sin (\Omega z) (y' \xh - x' \yh) ,
\end{equation}
from which we find the in-plane projection of the filament tangent configuration,
\begin{equation}
\label{eq: twist}
\tv_\perp(\rv) \simeq \partial_z \uv_\Omega= \Omega r \hat{\phi} .
\end{equation}
This expression for $\tv$---given in terms the current position, $\rv$, of the filament in cylindrical coordinates---describes a so-called ``double-twist" texture, well-studied in the context of liquid-crystal blue phases~\cite{wright_mermin}. Here, filament orientation rotates around the radial direction from the orientation, $\zh$, at the bundle center to a maximum tilt-angle, 
\begin{equation} 
\theta_{max} =\arctan(\Omega R) \simeq \Omega R,
\end{equation} 
at the outer boundary, where $R$ is the cylindrical radius of the bundle.

In this paper, we consider the rate of bundle twist, $\Omega$, to be quenched, say by the presence of strong, chiral inter-filament torques~\cite{grason_09} or, instead, by some external mechanical force.  According to helical symmetry described by eq. (\ref{eq: rot}), the filament positions, displacements and strains are described by the state of a given $xy$ plane at $z=0$, with the bundle at other $z$ described by uniform rotations of these configurations around the central axis.  In this case, the problem reduces to solving for $\uv(\xv)$ in the two-dimensional $z=0$ plane that minimizes the energy per unit length described by eq. (\ref{eq: elastic}), $E/L = \frac{1}{2} \int dA ( \lambda u_{kk}^2 + 2 \mu u_{ij}u_{ij} )$.  Given the composite displacement field of eq. (\ref{eq: rot}), the variation of the elastic energy with respect to components of $\uv$ gives
\begin{equation}
\label{eq: dE}
\frac{ \delta (E/L)}{\delta u_i} = -\partial_i \sigma_{ij} +\partial_z \big[ t_i \sigma_{ij} \big] ,
\end{equation}
where the stress tensor is simply
\begin{equation}
\label{eq: sigij}
\sigma_{ij} = \lambda u_{kk} \delta_{ij} + 2 \mu u_{ij} .
\end{equation}
The non-linear coupling between in-plane strain and out-of-plane filament tilt accounts for second contribution to the right-hand side of eq. (\ref{eq: dE}).  This represents a mechanical coupling of in-plane stress between different vertical layers of the bundle, which is absent from the linear theory of columnar order.  

Below we show that twist gives rise in-plane stresses, $\sigma_{ij}$, or order $(\Omega R)^2$.   It is straightforward to show from eqs. (\ref{eq: rot}) and (\ref{eq: twist}) that the factor of in-plane tilt, $t_i$ and $z$-derivative of contribute to $\partial_z \big[ t_i \sigma_{ij} \big]$ each an additional factor of $(\Omega R)^2$ relative to the terms $\partial_i \sigma_{ij}$ in (\ref{eq: dE}).   Hence, for the case that small twist, we approximate the Euler-Lagrange equations, $\delta E/\delta u_i=0$, by the order $(\Omega R)^2$ terms and neglect the higher-order corrections deriving from inter-plane coupling which contribute at order $(\Omega R)^4$.  Based on this approximation, we arrive at the condition for mechanical equilibrium from the elasticity theory of isotropic materials, namely,
\begin{equation}
\label{eq: divsig}
\partial_i \sigma_{ij} =0,
\end{equation}
along with the condition of vanishing normal stress at the boundary,
\begin{equation}
\hat{r}_i \sigma_{ij}(r=R) =0 .
\end{equation}

The problem of mechanical equilibrium in twisted bundles is one of 2D elasticity for which it is most convenient to solve $\sigma_{ij}$ in terms of an Airy stress function~\cite{landau_lifshitz}, $\chi$, related to the stress by,
\begin{equation}
\sigma_{ij} = \epsilon_{i k } \epsilon_{j \ell} \partial_k \partial_\ell \chi ,
\end{equation}
By construction, the stress derived from any function $\chi$  is divergence-free, satisfying mechanical equilibrium according to eq. (\ref{eq: divsig}).  However, not every $\chi$ corresponds to a physical configuration, which requires the existence of a displacement field $\uv(\xv)$.  That is, the solution for $\chi$ must be compatible with the definition of strain, $u_{ij}$.  

Similar to the well-known F\"oppl-von K\'arm\'an theory of thin sheets~\cite{landau_lifshitz}, we derive the compatibility condition for the Airy stress from the constitutive relationship between stress and strain,
\begin{equation}
\label{eq: uchi}
u_{ij} = \frac{  \epsilon_{i k } \epsilon_{j \ell} \partial_k \partial_\ell \chi }{ 2 \mu}- \frac{ \lambda \grad_\perp^2 \chi }{ 4 \mu (\lambda + \mu)} \delta_{ij} .
\end{equation}
We cast the conditions that eq. (\ref{eq: uchi}) can be solved to find a displacement field by taking the anti-symmetric derivatives of strain, $\epsilon_{i k} \epsilon_{j \ell} \partial_k \partial_\ell u_{ij}$, yielding the following compatibility relation for the Airy stress
\begin{equation}
\label{eq: Airy1}
\frac{1}{K_0} \grad_\perp^4 \chi = \frac{1}{2}  \epsilon_{i k} \epsilon_{j \ell} \partial_k \partial_\ell (\partial_i u_j +\partial_j u_i) -  \frac{1}{2}  \epsilon_{i k} \epsilon_{j \ell} \partial_k \partial_\ell t_i t_j ,
\end{equation}
where we have used the definition of stain in eq. (\ref{eq: ut}) and  $K_0 = 4 \mu(\lambda+ \mu)/(\lambda + 2 \mu)$, is the 2D Young's modulus of the array.  For single-valued strain functions, the first term on the right-hand side must vanish, but in the presence of topological defects in the cross section, these terms generate non-vanishing sources for Airy stress.  

In general, two types of topological defects contribute to the Airy stress of the two-dimensional array:  {\it disclinations}, associated with singular configurations of $\theta_6(\xv)$ the bond-angle field that points to the six-fold directions of lattice order; and {\it dislocations}, associated with singular configurations of $\uv(\xv)$ that lead to no far-field rotation of the lattice directions~\cite{nelson_defects}.  The bond-angle field describes local rotations of the lattice directions in the cross section, and hence, is related to the anti-symmetric, in-plane derivatives of the displacement,
\begin{equation}
\theta_6 = \frac{1}{2} \epsilon_{ij} \partial_i u_j .
\end{equation}
In a hexagonal lattice, a disclination indicates a point around which $\theta_6$ increases or decreases by an integer multiple of $2 \pi/ 6$,
\begin{equation}
\oint d {\bf \ell} \cdot \grad_\perp \theta_6 = s ,
\end{equation}
where $s = (2 \pi /6) n$ is the topological charge of the disclination.  From Stokes theorem we have the the areal density of disclinations, $s(\xv)$,
\begin{equation}
 s(\xv)
= \sum_\alpha s_\alpha \delta^{(2)} (\xv_\perp - \xv^\alpha_\perp) ,
\end{equation}
where the sum over $\alpha$ indicates a sum over multiple disclinations in the cross section.  Likewise, a dislocation is defined in terms of a closed loop integral around which $\uv$ changes by an integer multiple of the lattice spacing, $a$, along one of the six-fold directions,
\begin{equation}
\oint d {\bf \ell} \cdot \grad_\perp u_i = b_i,
\end{equation}
where $\bv$ is the Burger's vector.  The areal density of dislocations, ${\bf b}(\xv)$, is therefore, 
\begin{equation}
 {\bf b}(\xv) = \sum_\beta {\bf b}^\alpha_i \delta^{(2)} (\xv_\perp - \xv^\alpha_\perp) .
\end{equation}
By manipulating the displacement derivatives in the compatibility relation it is straightforward to show~\cite{nelson_seung} that disclinations and dislocations generate point-like and dipole-like sources, respectively for Airy stress,
\begin{equation}
 \frac{1}{2}  \epsilon_{i k} \epsilon_{j \ell} \partial_k \partial_\ell (\partial_i u_j +\partial_j u_i)  = s(\xv) - \grad_\perp \times {\bf b}(\xv) .
 \end{equation}
For the purposes of the present study, we focus on the case of disclinations only, ${\bf b} =0$, and solve exactly for the elastic energy of twisted bundles in the presence of arbitrary disclination configurations.  In a subsequent study~\cite{azadi_grason}, we take advantage of the fact that more complex topological defects, dislocations and grain boundaries, may be constructed from neutral configurations of multiple disclinations, whose elastic stresses, according to the compatibility relation, are superposable.  

In addition to the defect-induced sources for Airy stress, eq. (\ref{eq: Airy1}) shows that certain gradients of in-plane filament tilt generate a more homogenous source for $\chi$ due to the non-linear coupling of tilt and strain.  We denote these terms as the {\it intrinsic twist}, $K_T$, of the bundle, which can written as,
\begin{eqnarray}
\label{eq: KT} 
\nonumber
K_T &\!\!\equiv\!\!&\frac{1}{2}  \epsilon_{i k} \epsilon_{j \ell} \partial_k \partial_\ell t_i t_j \\ &\!\!=\!\!& \frac{1}{2}\grad_\perp \times \big[ (\grad_\perp \times \tv_\perp) \tv_\perp - (\tv_\perp \times \grad_\perp) \tv_\perp \big] .
\end{eqnarray} 
Indeed, by combining the disclination- and tilt-induced sources for Airy stress we arrive at the final form of compatibility relation for $\chi$,
\begin{equation}
\label{eq: Airy2}
\frac{1}{K_0} \grad_\perp^4 \chi = s(\xv) - K_T.
\end{equation}
When $K_T \neq 0$ somewhere in the cross section of the bundle, eq. (\ref{eq: Airy2}) tells us that there must be in-plane stress.  

Though $K_T$ is an unfamiliar geometric quantity, written in the form of eq. (\ref{eq: KT}), we see that it is a total derivative of gradients of $\tv$ in the $xy$ plane reminiscent of the so-called ``saddle-splay" of a vector field, $\nv_\perp$,
\begin{equation}
\label{eq: KG}
K_G = \frac{1}{2} \grad_\perp \cdot \big[ (\grad_\perp \cdot  \nv_\perp )\nv_\perp - (\nv_\perp \cdot \grad_\perp) \nv_\perp \big] ,
\end{equation}
a total derivative term in the Frank elastic energy of nematically-ordered materials~\cite{degennes_prost}.  The saddle-splay operator is most familiar when $\nv$ is the normal to a surface that is weakly-deflected from the $xy$ plane, say a membrane, in which case, $K_G$ is the Gaussian curvature of the surface~\cite{kamien_rmp_02}.  From this point of view, the intrinsic twist, $K_T$, plays precisely the role played by Gaussian curvature in the non-linear theory of thin elastic membranes.  In membranes, the in-plane strain, $u_{ij}^{(m)}$, couples to out-of-plane deflections~\cite{nelson_seung}, described by the height function, $h(\xv)$, according to
\begin{equation}
u_{ij}^{(m)} \simeq \frac{1}{2} \big( \partial_i u_j + \partial_j u_i + \partial_i h \partial_j h \big) ,
\end{equation}
where the $(m)$ in the subscript is used to distinguish the non-linear strain of membranes from the non-linear strain of filament arrays, eq. (\ref{eq: ut}).  The compatibility relation for the in-plane stress of membranes is known as the one of two F\"oppl-von K\'arm\'an relations, describing the mechanical equilibrium of thin plates~\cite{landau_lifshitz}.  This relation was generalized by Seung and Nelson~\cite{nelson_seung} to include the presence of topological defects in the crystalline order of membrane and is identical to eq. (\ref{eq: Airy2}), with $K_T$ replaced by the Gaussian curvature of the sheet, calculated from eq. (\ref{eq: KG}) and $\nv_\perp \simeq - \grad_\perp h$.   

The compatibility equation for in-plane stress encodes two fundamental properties unique to the non-linear theory of membranes.   First, it quantifies the in-plane stretching needed to deform intrinsically flat sheets into shapes corresponding to $K_G \neq 0$, in accordance with the renowned {\it Theorema Egregium} of Gauss~\cite{kamien_rmp_02}.  And second, it demonstrates the formal equivalence between the far-field in-plane stress distribution generated by localized regions of non-zero Gaussian curvature, and point-like, orientational defects in the crystalline order of membrane, 5- or 7-fold disclinations.  Due to this second property, unlike crystalline order on planar 2D surfaces, topological defects---disclinations in particular---are well known to be necessary in the groundstates of crystalline order on surfaces with non-zero curvature, as disclinations of the appropriate charge are able to screen the curvature-induced stresses in-plane.  An example of this frustration, important in structural studies of such diverse materials as fullerenes~\cite{smalley}, viruses~\cite{caspar_klug} and particle-stabilized emulsions~\cite{bausch_science}, is known alternately as the {\it Thomson problem}, which seeks to describe the lowest-energy arrangement of repulsive, point-like particles on the surface of a sphere~\cite{aste_weaire}.  

Though ordered filament bundles possess a unique geometry among 2D ordered materials, the mechanical frustration between in-plane stresses and out-of-plane geometry in these materials is remarkably similar to that predicted by the non-linear elastic theory of membranes.  In the case of filament bundles, the intrinsic twist plays the role play by Gaussian curvature in membranes, necessarily introducing distortions of the inter-filament spacing.  Given the striking similarity, it is tempting, perhaps, to identify $K_T$ with the Gaussian curvature of some surface or set of surfaces, say, to which filaments are normal.  Such an identification easily fails for any twisted configuration of filaments $\grad_\perp \times \tv_\perp \neq 0$, for which no such surfaces exist~\cite{degennes}, and yet $K_T$ is demonstrably non-zero.   For the case of super-helical twist considered here, we find,
\begin{equation}
\label{eq: KT2}
K_T (\Omega) =\frac{1}{2} \big( \partial_x^2 t_y^2 + \partial_y^2 t_x^2 -2 \partial_x \partial_y t_x t_y\big) = 3 \Omega^2 .
\end{equation} 
Thus, superhelical double-twist generates a homogeneous and negative source of Airy stress in the cross section.  In this sense, we find that the in-plane stresses induced by bundle twist are formally equivalent to those induced by {\it positive} Gaussian curvature in thin sheets.  Formally, we may think of the intrinsic twist, $K_T$, as a uniform density of negatively charged disclinations, continuously distributed throughout the bundle cross section.  Hence, as the bundle grows larger in radius, the integrated topological charge of this distribution grows as $(\Omega R)^2\simeq \theta_{max}^2$, demonstrating that the frustration between uniform filament spacing and bundle twist grows rapidly with bundle size.

Finally, though twist generates stresses that are formally equivalent to those in spherically-curved membranes, it is important to point out a key difference between the optimal packing of twisted bundles and the Thomson problem.  Due to their closed topology, the bond network of point packings covering the sphere are required to possess exactly 12 excess 5-fold disclinations in all configurations, a requirement of topology rather than a ground-state property~\cite{bowick_travesset_nelson}.  Due to the presence of a free boundary, the problem of a twisted, 2D-ordered filament bundle is more closely aligned with the problem of a finite-sized crystalline domain on a spherically-curved surface~\cite{giomi_bowick_prb_07}.  The presence of a free boundary in of twisted bundle or curved crystal allows the net number of disclinations (5- or 7-fold) to adjust according the energetic balance of disclinations needed to screen geometrically-induced stresses.  Furthermore, as we show below, the requirement that normal components of stress vanish at the boundary of the domain leads to a strong-position dependence of the energetic costs for disclinations in various positions in the bundle cross section.

\subsection{Elastic energy of disclinations in twisted bundles}

In this section we present the exact solution for the Airy stress of twisted, cylindrical bundles with multiple disclinations in the cross section, and derive the effective elastic energy depending only on $\Omega$ and the positions and topological charges of disclinations.  We seek a solution to the compatibility relation,
\begin{equation}
\frac{1}{K_0} \grad_\perp^4 \chi = \sum_\alpha s_\alpha \delta^{(2)} (\xv_\perp- \xv_\perp^\alpha)  - K_T.
\end{equation}
The elastic energies  of sufficiently twisted bundles show a clear preference for an excess of 5-fold disclinations, indicating that the groundstate packing of twisted bundles is complex and irregular.  

This basic result is clearly demonstrated by considering the stress of a twisted bundle possessing a single, centered disclination.  Denoting $\chi_T$ as the stress induced by intrinsic twist,
\begin{equation}
\label{eq: chiT}
K_0^{-1} \grad_\perp^4 \chi_T = - K_T
\end{equation}
we search for the form $\chi_T$ which satisfies stress-free boundary conditions.  In polar coordinates the components of stress are related to $\chi$ by,
\begin{equation}
\sigma_{rr} =r^{-2} \partial_\phi^2 \chi +r^{-1} \partial_r \chi ,
\end{equation}
\begin{equation}
\sigma_{\phi \phi} =  \partial_r^2 \chi,
\end{equation}
and
\begin{equation}
\sigma_{r \phi} = - \partial_r (r^{-1} \partial_\phi \chi).
\end{equation}
Hence, for the case where $\chi_T$ is axisymmetric, the conditions that $\sigma_{r \phi} = \sigma_{rr} =0$ at the boundary reduce to $R^{-1}\partial_r \chi_{r=R} = 0$.  The solution to eq. (\ref{eq: chiT}) subject to this boundary condition is
\begin{equation}
\chi_T = \frac{K_0}{64} \big(2 R^2 r^2 -r^4) K_T .
\end{equation}
A single disclination at $r=0$ generates a stress denoted by $\chi_{single}$, satisfying,
\begin{equation}
K_0^{-1} \grad_\perp^4 \chi_{single} =  s \delta^{(2)} (\rv) ,
\end{equation}
and the stress-condition at $r=R$.  The solution for $\chi_{single}$ for a centered disclination in a 2D circular cross section is known from ref. \cite{nelson_seung},
\begin{equation}
\chi_{single} = \frac{K_0 s}{8 \pi} \Big(\frac{r}{R}\Big)^2 \big[ \ln (r/R) -1/2 \big] .
\end{equation}
From the Airy stress functions we have the twist- and disclination-induced pressure fields,
\begin{equation}
\sigma_{kk}^T = \frac{K_0}{8} (R^2 - 2 r^2) K_T ,
\end{equation}
and
\begin{equation}
\sigma_{kk}^{single} = \frac{K_0 s}{4 \pi} \big[1+ 2 \ln (r/R) \big] .
\end{equation}
Notice that the intrinsic twist induces as a dilation at the core of the bundle, $\sigma_{kk}^T >0$ for $r < R/\sqrt{2}$, and compression at the periphery, $\sigma_{kk}^T <0$ for $r > R/\sqrt{2}$.  This trend is reversed for a positively charged disclination, such as a 5-fold defect, where $\sigma_{kk}^{single}<0$ for $r < R e^{-1/2}$ and $\sigma_{kk}^{single}>0$ for $r > R e^{-1/2}$.  Hence, the induced stresses of 5-fold disclinations have a tendency to neutralize, or screen, the twist-induced stresses that derive from the non-linear coupling between in-plane strain and filament tilt.  

As a result of this tendency, the elastic energy associated with bundle twist and disclinations are strongly coupled.   From the Airy stress, the elastic energy can be computed from $\frac{1}{2} \int dV \sigma_{ij} u_{ij}$, which using eqs. (\ref{eq: sigij}) and (\ref{eq: uchi}), can be manipulated to yield
\begin{multline}
E/L = \frac{1}{2}\int dA \Big\{ K_0^{-1} (\grad_\perp ^2 \chi)^2 \\ - \mu^{-1} \epsilon_{ik} \epsilon_{j \ell} \partial_k \partial_\ell \big( \partial_i \chi \partial_j \chi \big) \Big\} .
\end{multline}
The second term in the integrand above is equivalent to the total derivative, $\mu^{-1} \partial_k(\sigma_{k j} \partial_j \chi )$, whose contribution, therefore, vanishes due to the stress-free boundary condition.  Hence, the elastic energy derives directly from the in-plane pressure distribution, $\sigma_{kk} = \grad_\perp^2 \chi$,
\begin{multline}
\label{eq: Esingle}
E_{single} (\Omega) = \frac{\pi L }{K_0} \int dr~r(\sigma^T_{kk}+\sigma^{single}_{kk})^2 \\
  =  V K_0 \bigg(\frac{ 3 (\Omega R)^4}{128} + \frac{ s^2}{32 \pi^2 } - \frac{3 (\Omega R)^2 s}{ 64 \pi } \bigg) .
\end{multline}
The first term above represents the unscreened, non-linear elastic cost of bundle twist, whose $\Omega$- and $R$-dependence derives from the twist-induced strain, $u_{\phi \phi} \approx - (\Omega r)^2/2$.  The second term represents the self-energy of a single disclination, which can be understood in terms of the constant axial strain required to remove or add angular wedges to the lattice, $u_{\phi \phi} \approx s/(2 \pi)$.  The final term in eq. (\ref{eq: Esingle}) represents the interaction between twist and disclinations.  The sign of this term indicates a favorable interaction between twist and 5-fold disclinations, $s= + 2 \pi/6$, due to mutually screening of the far-field stresses generated by these distortions.  However, while the elastic cost of twist depends continuously on $\Omega$, disclination charge is discretized in units of $2 \pi/6$.  The minimal topological charge of 5-fold dislocations sets a critical value of bundle twist, 
\begin{equation}
(\Omega R)_c = \sqrt{ \frac{ 2}{9} } \simeq 0.47,
\end{equation}
above which the elastic screening by the disclination is sufficient to compensate for the defect self-energy cost.  According continuum theory, bundles twisted beyond this threshold are unstable to the incorporation of one or more 5-fold disclinations in the cross section of the filament lattice.

The calculation of the elastic energy of multi-disclination configurations follows a similar method to the case of a single, centered disclination.  However, when one or more defects is off-center, the stress is no longer axisymmetric, requiring a somewhat more complex form.  Though algebraically tedious, the solution to eq. (\ref{eq: Airy2}) proceeds by a standard multi-pole expansion of the disclination-induced stresses.  Ultimately, the infinite series representing defect interactions with twist-induced stress as well as defect-defect interactions for disclinations in arbitrary positions can be resummed exactly to yield the relative compact expressions in the effective theory for defects in twisted bundles described below.  As we know of no other published calculation of the exact disclination energy in cylindrical and circular crystals, we present the full details of this analysis in Appendix B.  We note that this theory may have other applications, for example, for multi-disclination configurations in curved domains of 2D crystalline membranes.  In the remainder of this section, we outline the solution and summarize the results.  

Consider a disclination at polar coordinates $(\rho, \phi=0)$ in the cross section of the bundle.  Following ref.~\cite{brown} we decompose the Airy stress of the defect into a {\it direct} term, $\chi_d$, describing the singular stress at the core, and an {\it induced} term, $\chi_i$, due to the interaction of the defect with the free boundary.  In the presence of a disclination of charge, $s$, the direct stress at a point $(r, \phi)$ in the cross section has the form
\begin{equation}
\chi_d = \frac{ K_0 s}{ 16 \pi} r'^2 \ln r' ,
\end{equation}
where, $r'^2 = r^2  + \rho^2 - 2 \rho r \cos \phi$ and $K_0^{-1} \grad_\perp^4 \chi_d = s \delta^{(2)}(\rv')$.  To determine the form of the induced stress, we consider the multipole expansion of $\chi_d$,
\begin{multline}
\label{eq: chid}
\chi_d =(\rho^2+r^2-2 r \rho \cos \phi) \\ \times \begin{cases} \ln \rho- \sum_{n=1}^\infty \frac{1}{n} \big(\frac{r}{\rho} \big)^n  \cos (n \phi) & r< \rho \\ \ln r- \sum_{n=1}^\infty \frac{1}{n} \big(\frac{\rho}{r} \big)^n  \cos (n \phi) &  r> \rho \end{cases} .
\end{multline} 
The induced stress satisfies, $\grad_\perp^4 \chi_i =0$ for $r<R$ and has the generic form
\begin{equation}
\label{eq: chii}
\chi_i = \sum_{n=0}^\infty \Big[ C_n r^n \cos ( n \phi) + D_n r^{n+2} \cos (n \phi) \Big],
\end{equation}
with $C_0=C_1= 0$.  The coefficients $C_n$ and $D_n$ are determined by boundary condition, $\sigma_{\phi r}= \sigma_{rr}=0$ at $r=R$.  In terms of the Airy stress, the radial stress components vanish at the boundary when~\cite{brown}
\begin{equation}
\label{eq: mitchell1}
\chi|_{r=R} = \kappa_1 \cos \phi + \kappa_2 ,
\end{equation}
and
\begin{equation}
\label{eq: mitchell2}
R ~ \partial_r \chi|_{r=R} = \kappa_1 \cos \phi .
\end{equation}
With the coefficients determined (see Appendix B), we evaluate the stress induced by the free boundary from,
\begin{eqnarray}\nonumber
\sigma_{kk}^i &\!\! =\!\!& \grad_\perp^2 \chi_i \\ \nonumber
&\!\! =\!\!& \sum_{n=0}^\infty 4(n+1) D_n r^n \cos (n \phi),
\end{eqnarray}
and from eq. (\ref{eq: sigin}) we have the total pressure distribution of the disclination,
\begin{multline}
\sigma_{kk}^{disc} = \frac{ K_0 s}{ 2 \pi} \Big\{ \ln (r'/R) +1 - \frac{\rho^2}{R^2} \\ 
- \Big( \frac{ \bar{\rho}}{\bar{r}'} \Big)^2 \frac{R^2 - \rho^2}{2R^2}  - \ln (\bar{r}' / \bar{\rho} ) \Big\}, 
\end{multline}
where $\bar{\rho} = R^2/\rho$, and $\bar{r}'^2 = \bar{\rho}^2 + r^2 - 2 r  \bar{\rho} \cos \phi$ measures the square distance to an ``image" defect of opposite charge outside of the bundle representing the screening of the elastic stress by the free boundary. 

As shown in Appendix B, the stress profile of multiple disclinations may be superposed along with the twist-induced stress, $\sigma_{kk}^T$, and the total elastic energy may be calculated by area integration of $\sigma^2_{kk}(2 K_0)$, yielding the effective energy quoted in ref. \cite{grason_prl_10}, which may be decomposed into 3 parts,
\begin{equation}
\label{eq: Etot}
E = E_{twist} +E_{disc}+E_{int} .
\end{equation}
Here, $E_{twist}/V = 3 K_0  (\Omega R)^4/128$ is the non-linear elastic cost of bundle twist derived in eq. (\ref{eq: Esingle}).  From eqs. (\ref{eq: Eself}) and (\ref{eq: Escreen}) we may combine the self-energy and twist-disclination interaction into a single disclination energy, $E_{disc}$,
\begin{equation}
E_{disc}/V = K_0 \sum_\alpha V^{(s_\alpha)}_{disc}(\Omega, \rho_\alpha),
\end{equation}
described in terms an effective, position-dependent potential for disclinations in twisted bundles,
\begin{equation}
\label{eq: Vdisc}
V^{(s_\alpha)}_{disc}(\Omega, \rho_\alpha)=  \frac{ s_\alpha}{32 \pi}  \Big[ \frac{ s_\alpha}{ \pi} - \frac{ 3 (\Omega R)^2}{2} \Big] \Big( 1- \frac{\rho_\alpha^2}{R^2} \Big)^2 .
\end{equation}
Finally, the interactions between disclinations are described by 
\begin{equation}
E_{int}/V = \frac{K_0}{2} \sum_{\alpha \neq \beta} s_\alpha V_{int} (\rv_\alpha, \rv_\beta) s_\beta,
\end{equation}
where
\begin{multline}
\label{eq: Vint}
V_{int} (\rv_\alpha, \rv_\beta) =\frac{1}{ 16 \pi^2} \bigg\{ \Big(1-\frac{ \rho_\alpha^2}{R^2} \Big)  \Big(1-\frac{ \rho_\beta^2}{R^2} \Big) \\ 
+\frac{ |\rv_\alpha - \rv_\beta|^2}{R^2} \ln \bigg[\frac{ |\rv_\alpha - \rv_\beta|^2}{ (R^2- \rho_\alpha^2) (R^2-\rho_\beta^2 )/R^2+ |\rv_\alpha - \rv_\beta|^2  } \bigg] \bigg\}.
\end{multline}
It is straightforward to check from eqs. (\ref{eq: Vdisc}) and (\ref{eq: Vint}) that both the elastic self-energy and the disclination interactions vanish when $\rho_\alpha \to R$, a signature of the stress free boundary condition.

The key result of this analysis is the mutual screening of the elastic stresses of twist and disclinations in the bundle cross section described by the effective, single disclination potential $V^{(s_\alpha)}_{disc}(\Omega, \rho_\alpha)$.  Both the self-energy disclinations and the elastic interactions between twist and disclinations---respectively, the terms proportional to $s_\alpha^2$ and $-s_\alpha (\Omega R)^2$ in eq. (\ref{eq: Vdisc})---are strongest for disclinations at the core of the bundle and decrease to zero as disclinations are pushed to the boundary of the bundle.  However, while the self-energy always prefers to expel disclinations to the boundary, for the interaction between disclinations and twist can either be attractive or repulsive, depending on the sign of $s_\alpha$.  As discussed above, 5-fold disclinations, with positive topological charge ($s_\alpha= + 2 \pi /6$), are attracted by the twist-induced stresses to the core of the bundle.  For small twist, $\Omega < \Omega_c$, the self-energy of defects dominates over the elastic screening of twist, and $V^{(s_\alpha)}_{disc}(\Omega, \rho_\alpha) > 0$ is repulsive, with a maximum for disclinations at the core, $\rho_\alpha = 0$.  However, for sufficiently twisted bundles, $\Omega > \Omega_c$, the net elastic energy of single disclinations is attractive, favoring 5-fold disclinations at the core the bundle.  This attractive regime indicates that the screening of twist-induced stress dominates the elastic self-energy of defects, and ultimately, suggests that the optimal, ground-state packing of filament bundles of fixed twist are irregular, including one or more topological defects in the cross section.  

\begin{figure}
\center \epsfig{file=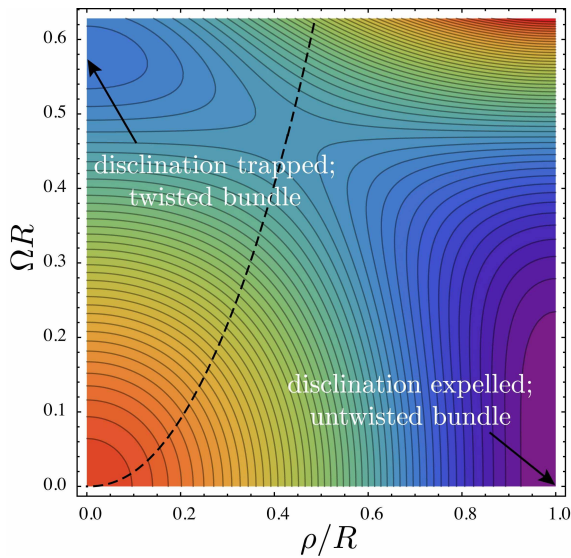, width=2.95in}\caption{The elastic energy landscape for a filament bundle possessing a single, 5-fold disclination in terms of defect radius, $\rho$, and twist of bundle, $\Omega$.  The color scale of the contour plot show lowest energies in purple and the highest energy values in red.  Two locally stable minima are labeled:  the global minimum, untwisted and defect free ($\rho = R, \Omega = 0$), and the meta-stable twisted state with trapped defect at core ($\rho =0 , \Omega = (\sqrt{3} R) ^{-1} $).  The dashed line shows a separatrix the divides the landscape into regions that flow along lines of force into each of the 2 minima.  Defects originating to the right of this boundary are expelled, while defects originating to the left are pulled towards the core of the twisted bundle. }
\label{fig: landscape}
\end{figure}

Before analyzing the structure of these multi-disclination groundstates in the next section, we briefly discuss the global stability of 5-fold disclinations in bundles that are free to relax both the radial position of the defect, as well as the bundle twist.  Fig. \ref{fig: landscape} shows a contour plot of the elastic energy, $E (\rho, \Omega) = E_{twist}+ E_{disc}$, of a bundle possessing a single 5-fold disclination as a function of radial position, $\rho$, and bundle twist, $(\Omega R)$.  In the absence of an external source for twist (i.e. imposed torque) or intrinsic source for twist (i.e. chiral filament interactions), the global minimum occurs when disclinations are expelled from untwisted bundles, $E (\rho=R, \Omega = 0)  =0$.  Fig. \ref{fig: landscape} shows three additional critical points for $0\leq \rho \leq R$ and $ \Omega R \geq 0$, where both the radial force on the disclination, $f_r=- \partial_r E$, and the twist-moment on the bundle, $M= - \partial_\Omega E$, vanish.  The point $\rho=\Omega = 0$ is a local maximum describing the self-energy of a centered-disclination in an untwisted bundle, $E_0 \equiv E(0,0)= K_0 V /288$.  Despite the absence an explicit energetic preference for bundle twist, we find that there exists a meta-stable twisted state for $\rho=0$ and $ \Omega_m = (\sqrt{3} R)^{-1}$, with a 5-fold disclination trapped at the bundle center for which $E(0, \Omega_m) = E_0/4$, the elastic cost of incorporating a disclination is significantly reduced due to the screening disclination stress by twist.  Finally, there is an unstable saddle-point separating the 2 minima at $\rho_s \equiv \rho =( 1 - \sqrt{6}/3 )^{1/2} R$ and $\Omega_s = \Omega_c$, for which $E(\rho_s, \Omega_s) = E_0/3$.  

The lateral motion of a disclination in 2D crystals requires a global repositioning of all lattice sites in the cross section, and therefore, we should expect the relaxation kinetics associated with defects under the forces described by $V_{disc}$ to be extremely slow.  Here we show that even in the absence of kinetic limitations on disclination motion, the non-linear elastic energy of filament bundles provides a thermodynamic mechanism to trap 5-fold disclinations in the cross section.   Assuming a crude model of overdamped kinetics of bundle twist, $\partial_t \Omega = \Gamma_\Omega M$, and defect motion, $\partial_t \rho = \Gamma_\rho f_r$, where $\Gamma_\Omega$ and $\Gamma_\rho$ are damping coefficients, we expect weakly twisted bundles to expel disclinations that originate outside of the bundle core as the system evolves to a lower energy.  However, when disclinations originate deep within the interior, the state of the bundle develops a finite twist, $\Omega_m = (\sqrt{3} R)^{-1}$, and the disclination is entrapped within the bundle center.  The boundary separating these two dynamical behaviors is shown as dashed line in Fig.~\ref{fig: landscape}.  While the entrapped disclination is ultimately thermodynamically unstable to expulsion from an untwisted bundle, there is an appreciable activation energy, $E_0/12$, proportional to the bundle volume, which must be overcome in order to remove the defect from the core of the twisted bundle.  For this reason, we argue that despite the extensive elastic energy needed to form them, due to the elastic screening provided by twist 5-fold disclinations formed at the core of two-dimensionally ordered filament bundles are especially long-lived, if not, permanent structural defects even in achiral materials, for which their is no intrinsic preference for bundle twist.

\section{Multi-Disclination Cross Sections of Twisted Bundles}

\label{sec: multi}

In this section we analyze the elastic-energy ground states of twisted-filament bundles based on the continuum theory of topological defects derived in Appendix B and summarized in eqs. (\ref{eq: Etot}) - (\ref{eq: Vint}).  These equations account only for the long-range elastic costs of deforming the array, and do not account, specifically, for the short-length scale effects captured by the core energy of disclinations, $E_{core}$.  In the following, however, we neglect this finite energy contributions in comparison to the elastic self-energy of disclinations that scales with $E_{self} \sim K_0 V$, and hence dominates in the limiting case of large bundles.  

\subsection{Equilibrium packings of twisted bundles}

In this section, we consider the minimal-energy configurations constructed from an integer number of 5-fold, $s = + 2 \pi/6$, disclinations.  According to the single-disclination elastic energy, eq. (\ref{eq: Vdisc}), for $(\Omega R) > (\Omega R)_c=\sqrt{2/9}$ 5-fold disclinations are attracted to the core of a twisted bundle.  While the non-linear stresses generated by twist attract disclinations to the bundle core, disclinations of equal sign repel one another at long range, countering the tendency to entrap an arbitrarily large number of defects in the core.  

Determining the ground state of disclinations in twisted bundles is, thus, not unlike the classical groundstate of long-range repulsive particles confined to finite, two-dimensional clusters~\cite{peeters_prb_94, partoens, peeters_pre_04}.  Such problems arise in a range of physical contexts including the Wigner-crystal states of electrons in mesoscopic particles~\cite{filinov}, 2D-confined colloidal crystals~\cite{bubeck_prl_99} and multi-vortex states of superconducting discs~\cite{peeters_prb_01} and quantum dots~\cite{saarikoski}.  The models previously studied differ from the present specifically in terms of the nature of both repulsive interactions and the effective confining potential.  Unlike the very long-range elastic interactions between disclinations described by eq. (\ref{eq: Vint}), the long-range potentials studied in previous cases are typically inverse power law for or logarithmic functions of separation.  Additionally, the elastic screening by the stress-free boundary in the present model leads to disclination interactions, eq. (\ref{eq: Vint}), that are weakened as either disclination approaches the boundary.  

\begin{figure}
\center \epsfig{file=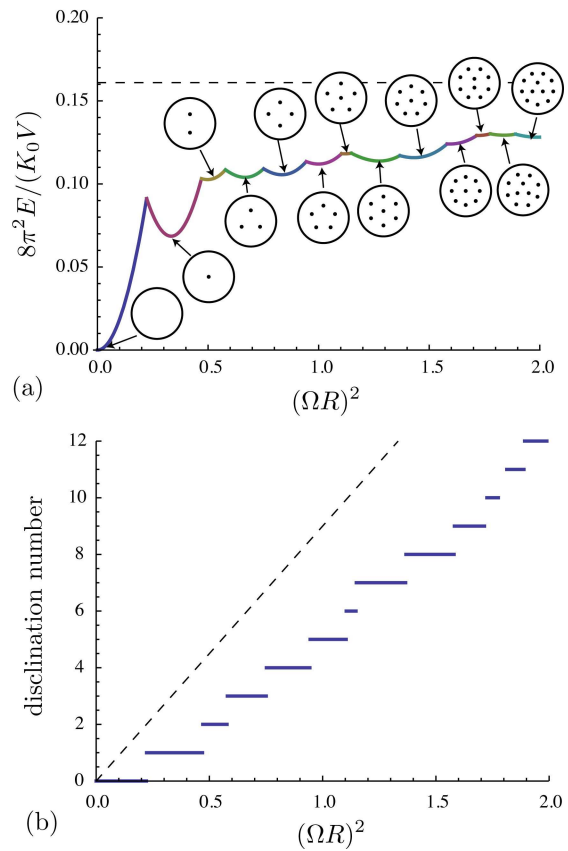, width=2.9in}\caption{In (a), we plot the elastic energy of ground state packings of twisted bundles predicted by continuum theory as a function of reduced twist.  The distinct sections of the curve correspond to ground states with a distinct number of disclinations in cross section.  Characteristic cross sections are shown for each disclination number, with points indicating the position of 5-fold disclinations in bundle.  The dashed horizontal line in (a) denotes the upper bound on the elastic energy provided by an infinite Wigner lattice of disclinations.  In (b), we plot the number of disclinations in the ground state in terms of bundle twist.  Here, the straight dashed line corresponds to the number of disclinations needed for perfect ``neutrality" of twist-induced stresses.  }
\label{fig: groundstate}
\end{figure}

The ground-state configurations of twisted bundles with only 5-fold disclinations were determined for twists in the range of $0< |\Omega R| \leq 0.45 \pi$ by direct comparison of the energies of defect configurations consisting of one or two disclination ``shells", a geometry common to few-particle groundstates of repulsive 2D clusters~\cite{peeters_prb_94}.  Each shell is composed of a $C_n$ symmetric arrangement of $n$ disclinations.  For a given twist, the elastic energy of the bundle cross section is minimized over the radii of the inner and outer shells, as well as the angular correlation between the shells.  The energy, structure and total disclination number of ground-state configurations are shown in Fig.~\ref{fig: groundstate}.  For bundle twist in the range $\sqrt{2/9} \leq \Omega R \leq 1.05$, the ground states possess a single shell of $1 \leq n \leq 5$ disclinations, while for higher twists, a second radial shell of disclinations is favored.  

While twist and its associated in-plane stresses increases continuously, the number of disclinations in the groundstate is quantized.  Figure \ref{fig: groundstate}b shows the step-wise increase in disclination number in increasingly twisted bundles, a roughly linear function of $(\Omega R)^2$.  The slope of the increase in disclination number is consistent, approximately, with the number of 5-fold disclinations required to ``neutralize" the effective topological charge of defect and twist-induced stresses.  From eq. (\ref{eq: Airy2}), we see that out-of-plane twist leads to an effective topological disclination charge per unit area, $-K_T = -3 \Omega^2$.  For a bundle of radius, $R$, the number of $s= + 2 \pi/6$ disclinations needed to neutralize this background charge can be estimated as,
\begin{equation}
n_{0}( \Omega R) = \frac{K_T A}{2 \pi /6} = 9 (\Omega R)^2 ,
\end{equation}
which is plotted as dashed line in Fig.~\ref{fig: groundstate}b for comparison to optimal disclination structures.   While the slope of this curve is roughly consistent with the defect number found by minimization of multi-disclination energy, notice that the number of disclinations predicted in the ground state is consistently less than $n_0 (\Omega R)$, suggesting that the neutrality condition is not precisely maintained for small bundles where much of twist-induced stresses are screened by the stress-free boundary condition.  

For large bundles, we expect the ground state predicted by the continuum theory of eq. (\ref{eq: Etoto}) to converge asymptotically a neutral, ``Wigner lattice" of disclinations.  The elastic energy of this configuration of the Wigner lattice can by calculated from the periodic solution to eq. (\ref{eq: Airy2}) for the in-plane pressure for an infinite plane,
\begin{equation}
\sigma_{kk}^{{\rm (WL) }} =\frac{K_T}{ K_0} \sum_{{\bf G} \neq 0} \frac{e^{i {\bf G} \cdot \xv}}{|{\bf G}|^2} ,
\end{equation}
where ${\bf G}$ are the reciprocal lattice vectors of a 2D lattice of areal density $K_T/(2 \pi /6)$~\cite{maradudin}.  Assuming the ground state of the Wigner lattice is a hexagonal array of disclinations, we calculate the elastic energy in the $\Omega R \to \infty$ limit from the lattice sum,
\begin{equation}
E_{{\rm (WL)}} /V = \frac{K_T^2}{2 K_0} \sum_{ {\bf G} \neq 0} \frac{1}{ |{\bf G}|^4} \simeq 0.00203 K_0^{-1} .
\end{equation}
Here, ${\bf G} = (2 \pi/d) \big[( 2 h+k) \hat{x} / \sqrt{3} + k \hat{y} \big]$ where $d = \sqrt{ 2 \pi /3^{3/2} K_T}$, the Wigner-lattice spacing of disclinations.  This constant value is shown as a dashed, horizontal line in Fig.~\ref{fig: groundstate}a.  As indicated by the series of multi-disclination groundstates found for $|\Omega R| \leq 0.45 \pi$, the appearance of 5-fold disclinations in the cross section of twisted bundles accounts for a crossover from a $(\Omega R)^4$ increase of the energy density to the ultimately twist-independent value of the infinite-radius, Wigner lattice state.  

It must be noted that in the forgoing analysis and derivation of the effect disclination energies,  in the equations of mechanical equilibrium, eq. (\ref{eq: divsig}), we have neglected terms that are higher-order in $(\Omega R)^2$ that derive from coupling of stresses in different vertical layers of the bundle.  Clearly, the inclusion of these effects must have a significant impact on the structure and energy of multi-disclination ground states for large values of reduced twist, when $\Omega R| \gg 1$.  However, the predicted value of  $(\Omega R)_c \simeq 0.47$ where the bundle becomes unstable to a single disclination, as well as the transition to multi-disclination ground states, fall well within the regime of twist where we expect the corrections to the approximate theory to be modest.  

\begin{figure*}
\center \epsfig{file=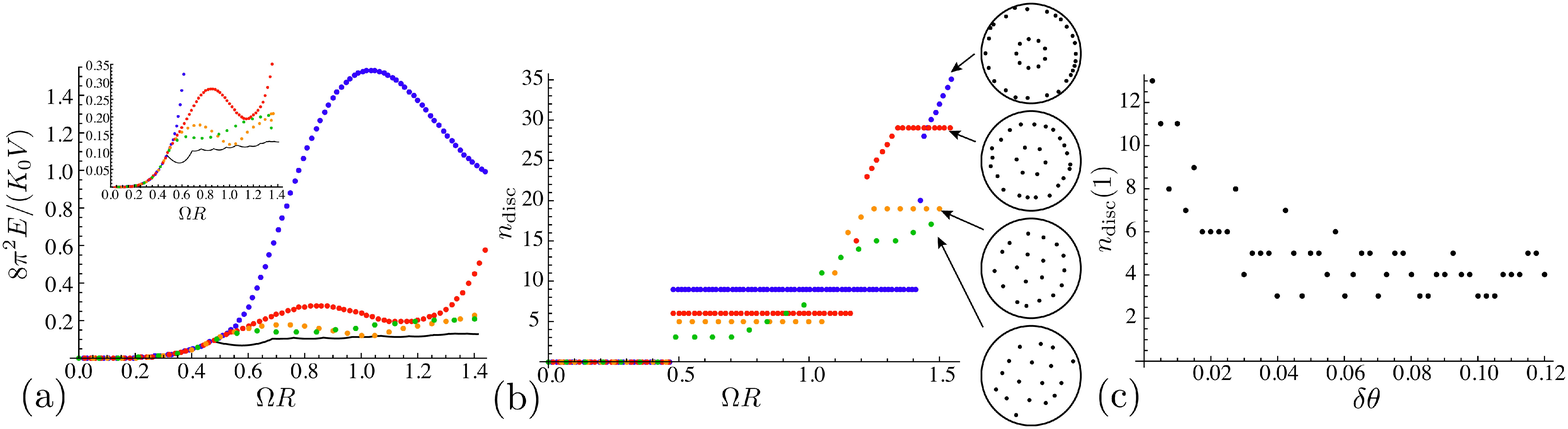, width=6.95in}\caption{The numerical results of the non-equilibrium model for radial growth of twisted bundles.  In (a), we show the elastic energy vs. twist for bundles grown with for different values of tilt increment:  $\delta \theta = 0.01$ (blue), 0.02 (red), 0.05 (orange) and 0.07 (green).  (b) shows the total number of disclinations in the bundle cross-section for these same values of $\delta \theta$, with cartoons depicted the position of 5-fold disclinations at the upper limit of $\Omega R$.  In (c), the number of disclinations in the first radial ``shell" (the first pile-up), is plotted as function of tilt increment. }
\label{fig: surface}
\end{figure*}

\subsection{Non-equilibrium disclination arrays in surface-assembled bundles}

The motion of disclination lines in twisted bundles of filaments is slow, and in many realistic scenarios the positions of disclinations that form in the cross section of twisted bundles are likely to be quenched during the assembly process.  The displacement of a disclination, say by a distance of one lattice spacing in the cross section, requires the absorption or emission of a dislocation, a neutral disclination pair, which in turn must diffuse from the disclination to the free boundary of the crystal~\cite{dewit, harris_scriven}.  Hence, the motion of a disclination over a distance of order $R$, requires a sequence of roughly $R/a$ dislocations to glide a distance of order $R/a$ lattice spacings.   We may crudely estimate the timescale for {\it disclination} motions within a 2D crystal grow at least as quickly $(R/a)^2 L$, where the factor of bundle length $L$ accounts for the fact that disclinations run along the full length the bundle.  All of this suggests that over the times scales during which the self-assembly process occurs, it is quite reasonable to expect that disclination motion is negligible, and therefore, the most probable state of even highly-twisted filaments assemblies will be markedly different from the equilibrium, multi-disclination states discussed in the previous section.

In this section we explore a simple model for the non-equilibrium formation of twisted-filament bundles in which the topological reorganization of the cross section is inhibited by the sluggish dynamics of disclinations.  In this model we assume that disclinations can only form at the free surface of a bundle as new filaments adhere to a partially-formed bundle.  Upon nucleation of disclinations within a surface layer of newly-adsorbed filaments, defects are then assumed to be frozen in position as further layers of absorbed filaments are assembled on the bundle.  

To study the cross sectional order within this model of surface assembly we construct the following algorithm where concentric radial layers of filaments are absorbed sequentially to a bundle of fixed rotation rate, $\Omega$.
\begin{enumerate}
\item Add radial filament layer of width $a$ to pre-existing bundle, increasing outer filament tilt by an increment, $\delta \theta = a \Omega$.
\item Consider addition of new disclinations at radial position, $R-a$, by minimizing eqs. (\ref{eq: Etot}) over new defect number and angular position.  Additional disclinations are subject to repulsive interactions within incipient layer, as well as with defects trapped within bundle from earlier stage of growth.
\item Energetically favorable defects are trapped within growing bundle and maintain their in-plane position upon growth of further surface layers.
\end{enumerate}
The minimization over disclination number and angular position is carried out numerically, and at each step in the radial-assembly process, the total number of disclinations, $n_{\rm disc}$, and total energy of the twisted-bundle cross section are determined.  

In Figure \ref{fig: surface} we show  results for the surface-assembly model of twisted bundles with 5-fold disclinations in the cross section.  These demonstrate that structures of the multi-disclination cross sections are both very distinct from the equilibrium cross sections of Fig. \ref{fig: groundstate}, as well as crucially sensitive to the tilt per filament layer, $\delta \theta = \Omega a$.  According to eq. (\ref{eq: Vdisc}), the stability of disclinations in twisted bundles is determined entirely by the reduced twist, $\Omega R$, the critical value of which is $\sqrt{2/9} \simeq  0.47$, is the same for all inter-filament spacings $a$, allowing us to correlate the magnitude of the tilt increment with the size of a bundle at the critical twist.  A smaller value of tilt per layer, $\Omega a$, implies that the bundle will be larger when it reaches the critical twist for stable disclinations.  This critical radius for disclinations is $R_*/a = 0.47/\delta \theta$, in units of filament spacing. 

Figure \ref{fig: surface}a shows the energy density of surface-assembled bundles for four values of tilt increment:  $\delta \theta = 0.01, 0.02, 0.05$ and $0.07$.  Surface-assembled bundles share the same $(\Omega R)^4$ energy-dependence below the critical value of twist,  and each show a characteristic maximum in $E/V$ due to the presence of 5-fold disclinations at higher twist.  Notably, these energies ``overshoot" the ground-state energy density above the critical value of $\Omega R$, and the extent of excess energy (relative to the groundstate) depends sensitively on tilt increment per layer, increasing dramatically as $\delta \theta$ is diminished.  In this model of assembly, larger bundles, measured in terms of $R_*/a$, screen twist-induced stresses less effectively than smaller bundles by the incorporation of 5-fold defects at their free boundaries.  This effect can be traced to the number of disclinations which are captured in the first few radial layers of growth beyond the critical size.  

The number of disclinations captured, $n_{\rm disc}$, at different values of $\Omega R$, is shown in Figure. \ref{fig: surface}b for the same four values of $\delta \theta$.  Each assembly exhibits an immediate ``pile-up" of 5-fold disclinations in the outer layers as soon as the bundle grows beyond the size necessary to stabilize a single disclination, $R_* = \sqrt{2/9} \Omega^{-1}$.  That is, just above the critical size, these assemblies jump from a defect-free state to assemblies with $n_{\rm disc} = 11, 6, 5$ and $3$ disclinations, respectively, for $\delta \theta = 0.01, 0.02, 0.05$ and $0.07$~\footnote{The ring of 5-fold defects in Fig. \ref{fig: surface}b is reminiscent of the disc-packing textures on spherical surfaces studied in M. Rubenstein and D. R. Nelson, Phys. Rev. B {\bf 28}, 6377 (1983).  Such packings were constructed by deterministic algorithm that grows locally-dense clusters radially outward from a given pole of the sphere.  While regions small compared to sphere radius retain hexagonal packing, upon reaching a critical lateral radius, these packings exhibit characteristic regions rich with excess 5-fold disclinations.}.  Hence, in comparison to energy-minimizing configurations of defects, when surface-assembled bundles reach the critical size, the assembly process tends to overcompensate for twist-induced stresses with a number of disclinations that far exceeds the number needed to neutralize the twist-induced stresses.  The number of excess 5-fold defects correlates directly with the excess elastic energy at large twist (Fig. \ref{fig: surface} a) as the defects are eventually incorporated in the core of the bundle where the elastic costs of inter-defect repulsions are more costly.  As shown in Fig. \ref{fig: surface} b the disclination pile-up tends to lead to the formation of radial bands of concentrated disclinations.  The larger the number of disclinations in the first pile-up, $n_{\rm disc} ( 1 )$, the higher the value of reduced twist, $\Omega R$, needed to produce the secondary band, as the elastic repulsion with the initial shell tends to stabilize a growing bundle to defects at further radial shells.

In Figure \ref{fig: surface} (c) we analyze the number of disclinations in the first radial ``shell", $n_{\rm disc} ( 1 )$, as a function of $\delta \theta$.  Here $n_{\rm disc} (1)$ is identified from the initial plateau value of $n_{\rm disc}$ beyond the critical twist.  Two trends are immediately apparent.  First, the value of  $n_{\rm disc} (  1 )$ is oscillatory with respect to $\delta \theta$.  Second, the mean number disclinations (the amplitude of the oscillatory envelope) in the first shell decreases with $\delta \theta$.  The oscillations in  $n_{\rm disc} ( 1)$ are simply due to the commensurability of a finite number of layers, $N_\ell$, of finite radial thickness, $a=\delta \theta / \Omega$, with the critical bundle size, $R_*$.  The number of layers of radial growth achieved when the bundle has grown just beyond the critical size is simply $N_* = 1+ {\rm int}  \big[\sqrt{2/9} / \delta \theta\big]$.  In the model of radial growth, a bundle will therefore be quenched just beyond the critical twist by an amount, $\delta (\Omega R) = N_* \delta \theta - \sqrt{2/9}$, at the point when the first shell of disclinations forms.  It can shown that due the integral nature of layer addition, this degree of overtwist oscillates with increasing $\delta \theta$ over a range 0 and $\delta \theta$ with a period roughly corresponding to $\delta \theta^2/\sqrt{2/9}$.

The decrease of the mean value $n_{\rm disc} (1)$ with $\delta \theta$ can be related to the weakening of repulsive interactions of disclinations near to the free boundary of the bundle.  According to the effective energy of disclinations in twisted bundles, eq. (\ref{eq: Etot}), $n_{\rm disc} (1)$ defects added near to the bundle boundary at $\rho = R -a$ leads to an elastic energy gain of roughly, $-n_{\rm disc} (1) K_0  \delta (\Omega R)a^2$.  Opposing the incorporation of multiple disclinations in the surface layer is the cost of inter-disclination repulsions.  From eq. (\ref{eq: Vint}), when defects approach the surface of the bundle such that $\rho_{\alpha} \to R$ (or $\rho_\beta \to R$) the inter-defect potential approaches zero as
\begin{equation}
\lim_{\rho_{\alpha} \to R} V_{int} (\rv_\alpha, \rv_\beta) =  \frac{  (R^2- \rho_\alpha^2)^2 (R^2-\rho_\beta^2 )^2 }{  32 \pi^2 R^6 |\rv_\alpha - \rv_\beta|^2 } .
\end{equation}
Assuming a multi-disclination shell with $n_{\rm disc} (1)$ defects spread evenly within a radial layer at $\rho = R - a$, we estimate the total cost of disclination repulsions per shell as roughly $n^2_{\rm disc} (1)K_0  V (a/R)^4$.  Although the favorable coupling between disclinations and twist vanishes as defects approach the bundle surface like $(a/R)^2$, the strength of disclination-disclination repulsion vanishes more rapidly as $(a/R)^4$, and thus, provides a weak resistance to the pile-up of disclinations at the surface layer of the bundle.   Optimizing the net elastic energy gain over the number of disclinations in the first shell we expect,
\begin{equation}
n_{\rm disc} (1) \approx  \frac{\delta (\Omega R)}{ (a/R_*)^2} \sim 1/\delta \theta.
\end{equation}
where the overtwist may be approximated by $\delta (\Omega R)\approx \delta \theta$.  This analysis shows, consistent with numerical results, that the number of disclinations in the first pile-up layer diverges as $\delta \theta \to 0$.  

The results of the simple kinetic model of bundle growth suggest that the efficiency of twist screening by elastically favorable 5-fold disclinations in non-equilibrium bundles is highly dependent on tilt per radial layer, $\delta \theta = \Omega a$.  Interactions between surface defects in bundles composed of relatively few filaments upon reaching the critical twist (i.e. $(R_*/a)^2 =\delta \theta^2$ is relatively small) tend stabilize the cross section against pile-up of excess disclinations.  While the excess defects trapped at the surface of relatively larger bundles ultimately lead to additional elastic costs to further bundle growth as the surface defects must ultimately be incorporated into the core of the bundle, where inter-disclination repulsive costs are greatest.  

\section{Discussion }

In this article, we studied the influence of out-of-plane geometry on the in-plane packing of twisted and two-dimensionally ordered filament bundles based on non-linear continuum elasticity theory.  Formally, this influence derives from the unique form of the non-linear 2D strain tensor describing deformation of columnar order, which couples inter-filament strains and tilting of filament backbones.  Analysis of the equations of mechanical equilibrium in filament bundles shows that helical twist generates stresses in the bundle cross section that are formally equivalent to those generated by positive (spherical) Gaussian curvature of the 2D elastic sheet.  In analogy to the problem of the curved, crystalline membranes, we also find here that these geometrically-induced stresses may be screened, in part, by 5-fold disclinations defects in the cross section of the bundle.  

We derived the effective theory of disclinations in the cross sections of twisted bundles and, based on this theory, studied the multi-disclination states of bundles for fixed reduced twist, $(\Omega R)$.  In equilibrium structures, disclinations appear above a critical twist, $( \Omega R)_c=\sqrt{2/9}$, and unlike the energy of the defect-free bundle, which grows unbounded with $\Omega^4$, the energy of the disclination-possessing groundstates remains below a finite, upper bound provided by the infinite-twist, Wigner lattice of disclinations.  In a simple model of non-equilibrium assembly of the twisted bundles, we show that the strong screening of disclination-disclination repulsions by the free boundary of the bundle leads to an instability to ``pile-up" excess disclinations at the surface of a bundle when the bundle grows to the critical value of reduced twist.  As the number of defects in these non-equilibrium packings far exceeds the number needed to neutralize the geometrically-induced stresses, in this model of bundle growth the presence of 5-fold disclinations is much less effective at reducing the in-plane elastic stresses that may act to limit the lateral growth of bundles during the assembly process.  

Underlying complex structure of the cross-sectional order in twisted bundles is the surprising and unusual connection highlighted by non-linear elasticity theory between crystalline order on spherically-curved surfaces and in twisted bundles.  We note here that the results of two similar models of chiral and biaxially-ordered materials anticipate this connection.  The first was demonstrated by Kamien for a chiral, liquid-crystal phase with both nematic and hexatic order, the so-called  $N+6$ phase~\cite{kamien_jphys_96}.  Here, the six-fold bond-orientational order is represented by a vector field, $\hat{{\bf N}}$, in a plane normal to the local nematic director, $\nv$.  The geometric relationship between $\hat{{\bf N}}$ and $\nv$---encoded in the Mermin-Ho relation---leads to the impossibility of uniform hexatic order for certain textures of the nematic director.  Indeed, for the double-twisted texture studied here, the intrinsic geometry of the nematic frustrates in-plane hexatic order in precisely the same way that curvature frustrates hexatic order formed on a spherical surfce.  Hence, in large double-twisted domains of the $N+6$ phase, just as in spherical hexatics, positively charged, 5-fold disclinations are favorable to screen geometrically-induced stresses.

Further evidence of the formal equivalence between frustrated order in twisted bundles and on spherical surfaces was demonstrated by Kl\'eman in studies of ideal packings of twisted filaments in an intrinsically curved, three-dimensional space, namely $S^3$~\cite{kleman_85_1, kleman_85_2}.  The results of these studies derive from the Hopf fibration, which decomposes $S^3$ into linked geodesic curves, great circles, that are each associated with a unique point on $S^2$~\cite{sadoc_frustration}.  The geometry of $S^3$ has the property that: 1) any two great circles are equidistant and twist helically along their length; and 2) the distance between them in $S^3$ is half of the separation of their associated points on $S^2$.  Hence, any equally-spaced arrangement of points on $S^2$ corresponds to a perfectly-packed (maximum density) and twisted configuration of curves, or filaments, in $S^3$.  In this case, the problem of point packings on $S^2$ has a one-to-one mapping onto a twisted fiber packing of $S^3$.  It is less clear at present, however, how this mapping may be used to study twisted filament packings in Euclidean space, as no projection from $S^3$ to $R^3$ preserves curve separations, leading to a second level of frustration associated with the flattening out of ideal packings of $S^3$ in addition to the frustration of point packings on $S^2$ already encoded in the fibration.  As with a certain other geometrically frustrated problems---notably liquid crystal blue phase textures and icosahedral order in amorphous materials~\cite{sadoc_frustration}--exploring the structure and energetics of ideal, twisted filament packings of $S^3$ projected to $R^3$ may ultimately provide a useful alternative avenue for studying chiral filament packing from a fundamentally geometrical point of view~\footnote{It is intriguing to note that both the theory of double-twisted $N+6$ order~\cite{kamien_jphys_96} and the model of ideal, twisted filament packing in $S^3$~\cite{kleman_85_1, kleman_85_2} find that both problems map on to the ordering on a sphere of curvature $K_G=\Omega^2$, that is 3 times smaller than the intrinsic twist result of eq. (\ref{eq: KT}) for twisted and 2D order filament packings in Euclidean space.}.

We conclude this article with a few remarks regarding the potential for observations of multi-disclination ground states in filamentous systems known to form twisted bundles or fibers.  According to continuum theory, in both equilibrium and non-equilibrium bundles the key parameter determining the appearance of 5-fold disclinations is the reduced twist, $\Omega R$, with a critical value $(\Omega R)_c= \sqrt{2/9} \simeq 0.47$.  This corresponds to a maximum tilt angle of the outer most filament of $\theta_c = \arctan[ (\Omega R)_c] \simeq 25.2^\circ$ with respect to the bundle axis.  This high degree of bundle twist has been demonstrated in coarse-grained simulation models of filament assembly~\cite{yue_jpcb_11}, and hence 5-fold defects are readily observed in the cross sections of simulated assemblies of chiral filaments~\cite{hagan_prl_10}.  However, this critical tilt exceeds the value observed for all if not most twisted, biofilament assemblies.  The collagen fibers that are observed to twist helically reach a maximum tilt angle of $15-16^\circ$~\cite{wess}.  The pitch and radius distributions of fibrin fibers have been well characterized {\it in vitro} by Weisel {\it et al.}~\cite{weisel_pnas_87}, showing a maximum radius of roughly 50 nm and a mean pitch 1.9 $\rm{\mu m}$, corresponding to an outer filament tilt of approximately $9^\circ$, falling well below the threshold at which disclinations become stable.  It is notable, however, that under certain conditions fibrin fibers are observed to grow to much larger radii, at which point the in-plane packing becomes highly-disrupted, leading to a significantly more diffuse, and possibly fractal, organization at the fiber periphery.  In light of the present study, we speculate that the appearance of disrupted in-plane packing of fibrin fibers beyond a critical size may be a remnant of the disclination pile-up that is predicted to occur in the non-equilibrium assembly model.  
 
\begin{acknowledgments}
The author would like to thank A. Azadi and I. Bruss for their careful review of this manuscript, as well as T. Lubensky for helpful discussions.  This work was supported by the NSF Career program under DMR Grant 09-55760.  
\end{acknowledgments}
\appendix
\section{Coarse-Grained Elastic Energy of a Microscopic Model of Self-Assembled Filament Bundles}

In this appendix we analyze a microscopic model of filamentous assemblies and deduce the form of the elastic energy via a coarse-graining of the microscopic degrees of freedom.  Our purpose is to demonstrate explicitly that the form of the elastic energy in eq. (\ref{eq: elastic}) described specifically in terms of the non-linear strain of eq. (\ref{eq: uij}) arises naturally for ordered filament bundles forming under the influence of generic and mutually-adhesive interactions.  For simplicity, we assume that interactions between a pair of filaments, labeled $\alpha$ and $\beta$, may be described by the following interaction energy,
\begin{equation}
\label{eq: Uab}
U_{\alpha \beta} =\frac{1}{2}\int ds_\alpha \int ds_\beta ~ v\big(|\rv_\alpha(s_\alpha) - \rv_\beta (s_\beta) | \big) ,
\end{equation}
where $\int ds_\alpha$ (or $\int ds_\beta$) denotes the integration over the arc-length of the central backbone of filament $\alpha$ (or $\beta$) which is described by $\rv_\alpha (s_\alpha)$ (or $\rv_\beta (s_\beta)$).  Here, $v(r)$ is the pair potential describing interactions between segments on different filaments, which we assume to be short-ranged and to be a function of $r^2$ for simplicity.  We carry out the second integration over $s_\beta$ for a fixed $s_\alpha$ by identifying $s'_\alpha$ as the arc-position on $\beta$ at the same vertical height as $\rv_\alpha (s_\alpha)$.  We may then approximate $\rv_\beta (s_\beta)$ to second order in $\delta s = s_\beta - s'_\alpha$ as,
\begin{equation}
\rv_\beta (s_\beta) = \rv_\beta (s'_\alpha) +  \tv_\beta(s'_\alpha) \delta s  + \frac{ \kappa_\beta (s'_\alpha)}{2} \nv_\beta(s'_\alpha) (\delta s)^2 + \ldots ,
\end{equation}
where $\tv_\beta(s'_\alpha)$,  $\nv_\beta(s'_\alpha) $ and $\kappa_\beta (s'_\alpha)$ are the tangent, normal and curvature of filament $\beta$ evaluated at $s'_\alpha$.  At each position $s_\alpha$ we denote the in-plane separation between the curves as $\bD_{\alpha \beta} (s_\alpha) = \rv_\alpha(s_\alpha) - \rv_\beta(s'_\alpha)$, while the square separation between two points at different vertical positions is approximately given by,
\begin{multline}
|\Delta \rv_{\alpha \beta}|^2 = \bD^2_{\alpha \beta} - 2 \bD_{\alpha \beta} \cdot \tv_\beta ~ \delta s \\ + \big(1 - \kappa_\beta \nv_\beta \cdot \bD_{\alpha \beta}\big) \delta s^2+ O(\delta s^3) ,
\end{multline}
where we have suppressed the $s_\alpha$ dependence for clarity.  In densely packed array, we expect filaments to be sufficiently straight on the scale of interfilament separation so that $\kappa_\beta \nv_\beta \cdot \bD_{\alpha \beta} \ll 1$ and we may ignore the curvature correction to $|\Delta \rv_{\alpha \beta}|^2$ above.  With this approximation, the separation between curves $\alpha$ and $\beta$ obtains a minimum for arc-separation,
\begin{equation}
\delta s_* \simeq \bD_{\alpha \beta} \cdot \tv_\beta, 
\end{equation}
which denotes the closest point to $\rv_\alpha(s_\alpha)$ on $\rv_\beta (s_\beta)$, or the {\it distance of closest approach}, 
\begin{equation}
\bD^2_{\alpha \beta, \perp} \equiv |\Delta \rv_{\alpha \beta}|_*^2 =  \bD^2_{\alpha \beta}  - ( \bD_{\alpha \beta} \cdot \tv_\beta)^2 .
\end{equation}
Hence, $\bD_{\alpha \beta, \perp}$ is perpendicular to $\tv_\beta$ at $s_\beta = s'_\alpha+\delta s_*$, and expanding around the distance of closest approach $\Delta s = \delta s - \delta s_*$, we have the separation between points on $\alpha$ and $\beta$,
\begin{equation}
|\Delta \rv_{\alpha \beta}|^2 \simeq \bD^2_{\alpha \beta, \perp}+ (\Delta s)^2 .
\end{equation}
Inserting this into eq. (\ref{eq: Uab}) and integrating over $ \Delta s$ we find that we may express the interaction between $\alpha$ and $\beta$ as the integral over $s_\alpha$ of a natural function of $\bD^2_{\alpha \beta, \perp}$, the local distance of closest approach,
\begin{equation}
\label{eq: Uab_local}
U_{\alpha \beta} = \int ds_\alpha V_{eff} ( \bD_{\alpha \beta, \perp} ) ,
\end{equation}
where taking the length of the filaments to be infinite
\begin{equation}
V_{eff}  ( \bD_{\perp} ) = \frac{ |\bD_{\alpha \beta, \perp}| }{2}\int_{-\infty}^{\infty} du~ v\big(|\bD_{\perp}|\sqrt{1+u^2} \big) .
\end{equation}
For example, for filaments whose segments interact via a Lennard-Jones potential, $v^{\rm{(LJ)}} (r) = \epsilon [(\sigma/r)^{12} - (\sigma/r)^6]$ it is straightforward to show that,
\begin{equation}
V^{{\rm(LJ) }}_{eff}   ( \bD_{\perp} ) = \frac{\epsilon}{2} \Big[ \frac{ \gamma_{12} \sigma^{12}}{|\bD_{\perp}|^{11}} - \frac{ \gamma_{6} \sigma^6}{|\bD_{\perp}|^5} \Big] ,
\end{equation}
where $\gamma_n = \int_{-\infty}^{\infty} du/(1+u^2)^{n/2}$.  

Provided that the interactions are sufficiently short-ranged, it is clear that pair-wise interactions between infinitely long filaments can always be decomposed, as in eq. (\ref{eq: Uab_local}), into the arc-length integral along a given curve $\alpha$ of an effective potential energy.  This effect potentail depends only on $\bD_{\alpha \beta, \perp} = \bD_{\alpha \beta}- \tv_\beta \cdot \bD_{\alpha \beta}$, the local nearest distance between $\alpha$ and $\beta$.  Assuming cohesive interactions between filaments, $V_{eff}  ( \bD_{\perp} )$, obtains a minimum for some value, $| \bD_{\perp} |= \Delta_0$.  Expanding around this separation we will have,
\begin{equation}
V_{eff}  ( \bD_{\perp} ) = -\epsilon + \frac{k}{2}(| \bD_{\perp} | - \Delta_0)^2 ,
\end{equation}
with $V_{eff}  ( \Delta_0 ) = - \epsilon$ and $k = d^2 V_{eff} /d \Delta_\perp^2$.  We assume that interactions are sufficiently short-ranged that we may truncate interactions beyond the nearest neighbors, and the ground state is a uniformly-spaced, hexagonal array of straight filaments.  Deformations from this state may be described by the energy,
\begin{equation}
\label{eq: discrete}
E = E_0 + \frac{k}{2} \int dz \sum_{\langle \alpha \beta \rangle} \frac{ d s_\alpha}{d z}  (| \bD_{\alpha \beta, \perp} | - \Delta_0)^2 ,
\end{equation}
where $E_0$ is the energy of the equally-space, reference configuration and the sum is carried out over nearest neighbor pairs in the hexagonal array.  Here, $d s_\alpha/ dz = (\tv_\alpha \cdot \zh)^{-1}$ due to the change in arc-length within a vertical layer of width, $dz$.  For a given filament pair that is initially separated by the nearest-neighbor vector $\Delta_0 \rvh_{\alpha \beta}$, we may write the separation in the distorted state in terms of two continuous functions, the displacement, $\uv(\xv)$, and the filament tangent, $\tv(\xv) \simeq \zh + \partial_z \uv$,
\begin{equation}
\bD_{\alpha \beta, \perp} \simeq \Delta_0 \rvh_{\alpha \beta}+ \Delta_0 ( \rvh_{\alpha \beta} \cdot \grad) \uv -  \Delta_0 (\rvh_{\alpha \beta}  \cdot \partial_z \uv ) \tv,
\end{equation}
where we have retained corrections of first order in derivatives of $\uv$, the in-plane displacement.  The change in the square-separation is straightforward to compute,
\begin{equation}
|\bD_{ \perp}|^2 - \Delta_0^2 \simeq  \Delta_0^2 \hat{r}_i \hat{r}_j \big( 2 \partial_i u_j +\partial_i \uv \cdot \partial_j \uv - \partial_z u_i \partial_z u_j \big) ,
\end{equation}
where we have suppressed the indices, $\alpha$ and $\beta$, on $\rvh_{\alpha \beta}$ for clarity.  Substituting into eq. (\ref{eq: discrete}) and using the fact that for a given lattice position, $\alpha$, the average over nearest neighbor vectors in the lattice yields,
\begin{equation}
\langle \hat{r}_i \hat{r}_j \hat{r}_k \hat{r}_\ell \rangle  = \frac{9}{8} \big( \delta_{ij} \delta_{k \ell} +  \delta_{ik} \delta_{j \ell}+ \delta_{i \ell } \delta_{j k} \big) ,
\end{equation}
and we can write the change of energy upon deformation as,
\begin{equation}
E - E_0 \simeq \frac{1}{2} \int dV \big\{ \lambda u_{kk}^2 + \mu u_{ij} u_{ij} \big\} .
\end{equation}
Here, the elastic constants are 
\begin{equation}
\lambda=\mu = \frac{9 k \rho_0}{4} ,
\end{equation}
where $\rho_0$ is the in-plane density of the reference configuration and the strain tensor,
\begin{equation}
u_{ij} = \frac{1}{2} \big( \partial_i u_j + \partial_i \uv \cdot \partial_j \uv - \partial_z u_i \partial_z u_j \big).
\end{equation}
Notice that $u_{ij}$ is ``mixed representation" strain~\cite{stenull_lubensky}.  This means that the $\uv$ is implicitly constructed as a function of the undeformed, reference state material coordinates for in-plane variables $x$ and $y$, a Langrangian description, while it is a function deformed positions for the vertical $z$ coordinate as in an Eulerian description~\cite{chaikin_lubensky}.

\section{Elastic energy of multi-disclination configurations by multipole expansion}

\subsection{Single disclination stress and self-energy}

In this section we present the solution of the Airy stress for a single disclination at a radial position, $\rho$, in the cross section of the twisted cylindrical bundle of radius $R$ by multipole expansion.  The infinite series that result from these expansions may be resummed, making use of the following series for $|z|<1$,
\begin{equation}
\label{eq: s1}
\sum_{n=1}^\infty \frac{z^n}{n} = - \ln (1- z),
\end{equation}
\begin{equation}
\label{eq: s2}
\sum_{n=1}^\infty \frac{z^n}{n+1} = -1-z^{-1} \ln (1- z),
\end{equation}
\begin{equation}
\label{eq: s4}
\sum_{n=1}^\infty \frac{z^n(n+1)}{n^2} = -\ln (1- z) + {\rm Li}_2 (z),
\end{equation}
\begin{equation}
\label{eq: s5}
\sum_{n=1}^\infty \frac{z^n}{n(n+1)} =1 +z^{-1} \ln (1- z) -\ln (1- z) ,
\end{equation}
\begin{equation}
\label{eq: s6}
\sum_{n=1}^\infty \frac{z^n}{n^2(n+1)} =-1 -z^{-1} \ln (1- z) +\ln (1- z) + {\rm Li}_2 (z),
\end{equation}
and
\begin{equation}
\label{eq: s7}
\sum_{n=2}^\infty \frac{z^n}{n^2(n-1)} =2 z +\ln (1- z) -z\ln (1- z) - {\rm Li}_2 (z),
\end{equation}
where ${\rm Li}_2 (z)=\sum_{n=1}^\infty = z^n /n^2$ is a polylogarithm.

The coefficients of the induced stress, $\chi_i$, are chosen to satisfy the stress free conditions, eqs. (\ref{eq: mitchell1}) and (\ref{eq: mitchell2}).  From eq. (\ref{eq: chid}) we have the coefficients of the multi-pole expansion of the direct stress, 
\begin{equation}
\chi_d = \frac{K_0 s}{8 \pi} \times \begin{cases}  \sum_{n=0}^\infty \chi_{d,<}^{(n)} \cos (n \phi) & r< \rho \\ \sum_{n=0}^\infty \chi_{d,>}^{(n)} \cos (n \phi) & r> \rho \end{cases}, 
\end{equation}
where,
\begin{equation}
\chi^{(0)}_{d, <}= (\rho^2 + r^2) \ln \rho + r^2 ,
\end{equation}
\begin{equation}
\chi^{(1)}_{d, <}= -2 \rho r \ln \rho -\rho^2 - \frac{ r^3}{2 \rho} ,
\end{equation}
\begin{equation}
\chi^{(n\geq2)}_{d, <}= 2 \Big(\frac{r}{\rho} \Big)^n \Big[\frac{\rho^2}{n(n-1)} - \frac{r^2}{n(n+1)} \Big] ,
\end{equation}
and
\begin{equation}
\chi^{(0)}_{d, >}= (\rho^2 + r^2) \ln r + \rho^2 ,
\end{equation}
\begin{equation}
\chi^{(1)}_{d, >}= -2 \rho r \ln r -r^2 - \frac{ \rho^3}{2 r} ,
\end{equation}
\begin{equation}
\chi^{(n\geq2)}_{d, >}= 2 \Big(\frac{\rho}{r} \Big)^n \Big[\frac{r^2}{n(n-1)} - \frac{\rho^2}{n(n+1)} \Big] .
\end{equation}

For the case of stress-free boundaries, the elastic energy only depends on $\sigma_{kk} = \grad_\perp^2 \chi$, and hence, the terms in eq. (\ref{eq: chii}) proportional to the coefficients $C_n$ do not contribute to the energy.  Defining $D_n \equiv (K_0 s/ 8 \pi) d_n$, from $R ~ \partial_r \chi|_{r=R} = \kappa_1 \cos \phi $ we find that,
\begin{equation}
d_0=- \ln R - \frac{1}{2} - \frac{\rho^2}{2 R^2} .
\end{equation}
From $\chi|_{r=R} - R ~ \partial_r \chi|_{r=R} = \kappa_2$ we find,
\begin{equation}
d_1 =\frac{\rho}{R^2} -  \frac{\rho^3}{2 R^4} .
\end{equation}
Finally, the as all terms proportional to $\cos (n \phi)$ in $\chi$ and $\partial_r \chi$ vanish at $r=R$, we find the condition 
\begin{equation}
R ~ \partial_r \chi_{d,>}^{(n)}|_{r=R} - n \chi_{d, >}^{(n)}|_{r=R} +2 d_n R^{n+2} =0 , \ (n\geq 2),
\end{equation}
which is satisfied by
\begin{equation}
d_{n\geq 2} = \frac{1}{R^{n+2}} \Big( \frac{ \rho}{R} \Big)^n \Big[ \frac{R^2}{n} - \frac{ \rho^2}{n+1} \Big] .
\end{equation}
 From these coefficients, we may compute the induced pressure, $\sigma_{kk}^{i} = \grad_\perp^2 \chi_i$,
\begin{eqnarray}
\label{eq: sigin}
\nonumber
\sigma_{kk}^{i} &\!\!= \!\! &\frac{ K_0 s}{ 8 \pi} \sum_{n=0}^\infty 4(n+1) d_n r^n \cos(n \phi) \\ \nonumber &\!\!=\!\!&  \frac{ K_0 s}{ 8\pi}  \bigg\{ -4\ln R -2\frac{R^2+ \rho^2}{R^2} \\ \nonumber &&+ 4
{\rm Re} \bigg[ \sum_{n=1} \Big( \frac{r \rho}{R^2} e^{ i \phi} \Big)^n \Big( \frac{R^2 - \rho^2}{R^2} + \frac{1}{n} \Big) \bigg] \bigg\} \\ \nonumber &\!\!=\!\!&  
 \frac{ K_0 s}{ 8 \pi}  \Big\{ -4\ln R -4 \frac{\rho^2}{R^2}   \\ && - 2 \Big( \frac{ \bar{\rho}}{\bar{r}'} \Big)^2 \frac{R^2 - \rho^2}{R^2}  - 2 \ln (\bar{r}'^2 / \bar{\rho}^2 ) \Big\},
\end{eqnarray}
where $\bar{\rho} = R^2/\rho$ and $\bar{r}'^2 = \bar{\rho}^2 + r^2 - 2 r  \bar{\rho} \cos \phi$, and we have used eq. (\ref{eq: s1}) and the geometric series.  Note the appearance of a singularity at $\bar{r}' = 0$, which indicates the presence of an ``image" defect outside of the bundle boundary, at a distance $\bar{\rho} > R$ from the center of the bundle, which screens the normal stresses at the free boundary.  

The next task is to use the disclination-induced stress to calculated the elastic self-energy of disclinations and the elastic interaction between defect- and twist-induced stresses.  Due to the orthogonality of multipole terms with respect to $\phi$ integration, to perform area integration of the elastic energy it is most convenient to work with a full expansion of the defect induced pressure, $\sigma_{kk}^{disc} = \sigma_{kk}^d + \sigma_{kk}^i$,
\begin{equation}
\label{eq: sig1}
\sigma_{kk}^{disc} = \frac{ K_0 s}{ 8 \pi} \times \begin{cases} \sum_{n=0}^\infty \sigma^{(n)}_< \cos (n \phi) & r< \rho \\  \sum_{n=0}^\infty \sigma^{(n)}_> \cos (n \phi) & r> \rho \end{cases}, 
\end{equation}
where
\begin{equation}
\label{eq: sig2}
\sigma^{(0)}_< = 4 \ln (\rho/R) + 2 - 2 \frac{ \rho^2}{R^2} ,
\end{equation}
\begin{multline}
\sigma^{(n\geq1)}_< = 4\Big[ -\frac{1}{n} \Big(\frac{r}{\rho}\Big)^n + \frac{(n+1)}{n} \Big(\frac{ r \rho }{R^2} \Big)^n \\ - \Big(\frac{ r \rho }{R^2} \Big)^n \Big( \frac{ \rho}{R} \Big)^2 \Big] ,
\end{multline}
and 
\begin{equation}
\label{eq: sig3}
\sigma^{(0)}_>= 4 \ln (r/R) + 2 - 2 \frac{ \rho^2}{R^2} ,
\end{equation}
\begin{multline}
\label{eq: sig4}
\sigma^{(n\geq1)}_> = 4\Big[ -\frac{1}{n} \Big(\frac{\rho}{r}\Big)^n + \frac{(n+1)}{n} \Big(\frac{ r \rho }{R^2} \Big)^n \\ - \Big(\frac{ r \rho }{R^2} \Big)^n \Big( \frac{ \rho}{R} \Big)^2 \Big] .
\end{multline}
Defining 
\begin{equation}
E^{(n)}_< =  \int_0^\rho dr~r \big(\sigma^{(n)}_< (r) \big)^2 ,
\end{equation}
and
\begin{equation} 
E^{(n)}_> =  \int_\rho^R dr~r \big(\sigma^{(n)}_> (r) \big)^2 ,
\end{equation}
we may write the elastic self-energy of the disclination as
\begin{equation}
\label{eq: Eself}
E_{self}/L = \frac{ K_0 s^2}{ 64 \pi^2} \Big[ E^{(0)}_< + E^{(0)}_> + \frac{1}{2} \sum_{n=1}^\infty E^{(n)}_< + E^{(n)}_> \Big] .
\end{equation}
Evaluation of the radial integrals is straightforward.  For $n=0$ terms we have
\begin{eqnarray}
\nonumber
E^{(0)}_< &\!\!=\!\!& 16 \int_0^\rho dr ~r\Big[ \ln (\rho/R) +\frac{1}{2} - \frac{1}{2} \Big(\frac{ \rho}{R} \Big)^2\Big]^2 \\
&\!\!=\!\!& 8 \rho^2 \Big[ \ln (\rho/R) +\frac{1}{2} - \frac{1}{2} \Big(\frac{ \rho}{R} \Big)^2\Big]^2,
\end{eqnarray}
\begin{eqnarray}
\nonumber
E^{(0)}_
>&\!\!=\!\!& 16 \int_\rho^R dr ~r\Big[ \ln (r/R) +\frac{1}{2} - \frac{1}{2} \Big(\frac{ \rho}{R} \Big)^2\Big]^2 \\ 
&\!\!=\!\!& 2\frac{\rho^4}{R^2}+2R^2- 2 \rho^2 - 8\rho^2 \Big[ \ln (\rho/R) -\frac{1}{2} \Big(\frac{\rho}{ R}\Big)^2 \Big]^2
\end{eqnarray}
and
\begin{equation}
\label{eq: E0}
E^{(0)}_<+E^{(0)}_> = 8 \rho^2 \ln (\rho/R) + 2 R^2 - 2 \frac{ \rho^4}{R^2} .
\end{equation}
For $n\geq1$ and $E^{(n)}_<$ we have,
\begin{eqnarray}
\nonumber
E^{(n\geq1)}_< &\!\!=\!\!& 16 \int_0^\rho dr ~r^{2n+1}\Big[ -\frac{1}{n \rho^n} + \frac{(n+1)}{n} \Big(\frac{\rho}{R^2}\Big)^{n} - \frac{ \rho^{n+2}}{ R^{2n+2}} \Big]^2 \\
&\!\!=\!\!& \frac{8 \rho^2}{(n+1)}\Big[ -\frac{1}{n } + \frac{(n+1)}{n} \Big(\frac{\rho}{R}\Big)^{2n} - \Big(\frac{ \rho}{ R} \Big)^{2n+2} \Big]^2  .
\end{eqnarray}
For  $E^{(1)}_>$ 
\begin{eqnarray}
\nonumber
E^{(1)}_> &\!\!=\!\!& 16 \int_0^\rho dr ~r\Big\{ \Big(\frac{\rho}{r} \Big)^2 -2 \Big(\frac{\rho}{R} \Big)^2  \Big[2- \Big(\frac{\rho}{R} \Big)^2 \Big]  \\ \nonumber && + \frac{ r^{2} \rho^{2} }{ R^{4}}  \Big[2- \Big(\frac{\rho}{R}\Big)^2 \Big]^2 \Big\}  \\ \nonumber
&\!\!=\!\!& 8 \Big\{ 2 \rho^2 \ln (R/\rho) - 2(R^2 - \rho^2)\Big(\frac{\rho}{R} \Big)^2 \Big[2- \Big(\frac{\rho}{R} \Big)^2 \Big] \\ && +\frac{(R^4-\rho^4)}{ 2 R^2} \Big(\frac{\rho}{R} \Big)^2 \Big[2- \Big(\frac{\rho}{R} \Big)^2 \Big]^2 \Big\}
\end{eqnarray} 
and for $n \geq 2$ and $E^{(n)}_>$ we have
\begin{eqnarray}
\nonumber
E^{(n\geq2)}_> &\!\!=\!\!& 16 \int_0^\rho dr ~r\Big\{\frac{1}{n^2} \Big(\frac{\rho}{r} \Big)^{2n} -\frac{2}{n} \Big(\frac{\rho}{R} \Big)^{2n } \Big[\frac{(n+1)}{n}- \Big(\frac{\rho}{R} \Big)^2 \Big]  \\ \nonumber && + \frac{ r^{2n} \rho^{2n} }{ R^{4n}}  \Big[\frac{(n+1)}{n}- \Big(\frac{\rho}{R} \Big)^2 \Big]^2 \Big\}  \\ \nonumber
&\!\!=\!\!& 8 \Big\{\frac{1}{n^2(n-1)} \Big[ \rho^2 -R^2 \Big(\frac{\rho}{R} \Big)^{2n} \Big] \\ \nonumber &&  - \frac{2}{n}(R^2 - \rho^2)\Big(\frac{\rho}{R} \Big)^{2n} \Big[\frac{(n+1)}{n}- \Big(\frac{\rho}{R} \Big)^2 \Big] \\ && +\frac{(R^{2n+2}-\rho^{2n +2})}{R^{2n} (n+1)} \Big(\frac{\rho}{R} \Big)^{2n} \Big[\frac{(n+1)}{n}- \Big(\frac{\rho}{R} \Big)^2 \Big]^2 \Big\} .
\end{eqnarray} 
Using these expressions and eqs. (\ref{eq: s1}) -(\ref{eq: s7}) we find,
\begin{equation}
\frac{1}{2} \sum_{n=1}^{\infty} E^{(n)}_<+E^{(n)}_> = 4 \rho^2 \Big(\frac{\rho^2}{R^2} -1 \Big) + 8 \rho^2 \ln (R/\rho) .
\end{equation}
Using this and eq. (\ref{eq: E0}) in eq. (\ref{eq: Eself}) we have,
\begin{equation}
\label{eq: Eself}
E_{self}/L = \frac{ K_0 s^2}{ 32 \pi} R^2 \Big(1- \frac{ \rho^2}{R^2} \Big)^2 .
\end{equation}
Note that both the energy and force on the disclination due to the self-interaction vanish at the boundary as $\rho \to R$.  

Because $\sigma^T_{kk}$ contains only an monopole ($n=0$) term, it is relatively straightforward to compute the mutual screening of the twist- and defect-induced stresses,
\begin{eqnarray}
\nonumber
\label{eq: Escreen}
E_{screen}/L &\!\! =\!\! & \frac{ s}{ 8 } \int_0^R dr~r~ \sigma^{(0)}(r)  \frac{ K_0 }{8}(R^2-2r^2) K_T \\
 &\!\! =\!\! & -\frac{ K_0 s}{ 128} K_T R^4 \Big(1-\frac{\rho^2}{R^2} \Big)^2 .
\end{eqnarray}

\subsection{Disclination interactions}

Here, we use the multipole expansion of the single disclination stress profile derived above, eqs. (\ref{eq: sig1}) -(\ref{eq: sig4}), to calculate the elastic energy of interaction between pairs of disclinations in the cylindrical cross section of the two-dimensionally order material.  We consider two disclinations with respective topological charge, $s_1$ and $s_2$, located at polar coordinates $(\rho_1, 0)$ and $(\rho_2, \phi)$, and without loss of generality, assume $\rho_2 >\rho_1$.  The elastic energy is computed from the following area integral,
\begin{equation}
E_{12}/L= K_0^{-1} \int dA ( \sigma_{kk}^{d1} + \sigma_{kk}^{i1})  ( \sigma_{kk}^{d2} + \sigma_{kk}^{i2}),
\end{equation}
where $\sigma_{kk}^{d\alpha}$ and $\sigma_{kk}^{i\alpha}$ denotes the direct and induced stresses generated by the $\alpha$th defect.  Expanding the product in the integrand above, we consider each contribution in turn, beginning with the induced-induced terms,
\begin{eqnarray} 
\label{eq: Eii}
\nonumber
E_{ii} &\!\!=\!\! & \Big( \frac{ K^2_0 s_1 s_2 }{ (8 \pi)^2} \Big)^{-1} \int_0^R dr ~r ~ \sigma_{kk}^{i1} (r) \sigma_{kk}^{i2} (r) \\ \nonumber
 &\!\!=\!\!& 16 \int_0^R dr ~r \Big\{ \Big( \ln R + \frac{1}{2} + \frac{ \rho_1^2}{2 R^2} \Big) \Big(\ln R + \frac{1}{2}  + \frac{ \rho_2^2}{2 R^2}\Big) \\ \nonumber && + \frac{1}{2} \sum_{n=1}^\infty \Big( \frac{ r^2 \rho_1 \rho_2}{R^4} \Big)^n \Big[\frac{(n+1)}{n} - \Big(\frac{\rho_1}{R} \Big)^2 \Big] \\ \nonumber && \times \Big[\frac{(n+1)}{n} - \Big(\frac{\rho_2}{R} \Big)^2 \Big] \cos (n \phi) \Big\} \\ \nonumber &\!\!=\!\!& 8 R^2\Big\{ \Big( \ln R + \frac{1}{2} + \frac{ \rho_1^2}{2 R^2} \Big) \Big(\ln R + \frac{1}{2}  + \frac{ \rho_2^2}{2 R^2}\Big) \\ 
 \nonumber && +\frac{1}{2} \sum_{n=1}^\infty\frac{1}{n+1}\Big( \frac{  \rho_1 \rho_2}{R^2} \Big)^n \Big[\frac{(n+1)}{n} - \Big(\frac{\rho_1}{R} \Big)^2 \Big]  \\ \nonumber && \times \Big[\frac{(n+1)}{n} - \Big(\frac{\rho_2}{R} \Big)^2 \Big] \cos (n \phi) \Big\} \\ \nonumber 
 &\!\!=\!\!& 8 R^2\bigg\{ \Big( \ln R + \frac{1}{2} + \frac{ \rho_1^2}{2 R^2} \Big) \Big(\ln R + \frac{1}{2}  + \frac{ \rho_2^2}{2 R^2}\Big) \\ 
 \nonumber && +\frac{1}{2} {\rm Re} \Big[ - \ln\Big( 1- \frac{ \rho_1 \rho_2}{ R^2} e^{i \phi}\Big) + {\rm Li}_2 \Big( \frac{ \rho_1 \rho_2}{ R^2}  e^{i \phi}\Big) \\ \nonumber && + \frac{ (\rho_1^2+ \rho_2^2)}{R^2}  \ln\Big( 1- \frac{ \rho_1 \rho_2}{ R^2} e^{i \phi}\Big)  -\frac{ \rho_1^2 \rho_2^2}{R^4}  \\ && - \frac{\rho_1 \rho_2}{R^2}e^{- i \phi}  \ln\Big( 1- \frac{ \rho_1 \rho_2}{ R^2} e^{i \phi}\Big)   \Big]\bigg\} .
\end{eqnarray}
Next, we consider the direct-direct terms,
\begin{eqnarray}
\nonumber
E_{dd} &\!\! = \!\!& \Big( \frac{ K^2_0 s_1 s_2 }{ (8 \pi)^2} \Big)^{-1} \int_0^R dr~r~ \sigma_{kk}^{d1} (r) \sigma_{kk}^{d2} (r) \\ \nonumber
 &\!\!=\!\!& 16 \bigg\{\int_0^{\rho_1} dr ~r \Big[ \big(\ln \rho_1 + 1 \big)\big(\ln \rho_2 +1\big) \\ \nonumber && \ \ + \frac{1}{2}\sum_{n=1}^\infty \frac{1}{n^2} \Big( \frac{ r^2}{ \rho_1 \rho_2 } \Big)^n \cos (n \phi) \Big] \\ \nonumber && + \int_{\rho_1}^{\rho_2}dr~r\Big[ \big(\ln r + 1 \big) \big(\ln \rho_2 +1 \big)   \\ \nonumber && \ \ + \frac{1}{2}\sum_{n=1}^\infty \frac{1}{n^2} \Big( \frac{ \rho_1}{ \rho_2 } \Big)^n \cos (n \phi) \Big] \\ \nonumber && + \int_{\rho_2}^{R}dr~r\Big[ \big(\ln r + 1 \big) \big(\ln r +1 \big)  \\ \nonumber && \ \  + \frac{1}{2}\sum_{n=1}^\infty \frac{1}{n^2} \Big( \frac{ \rho_1\rho_2}{ r^2 } \Big)^n \cos (n \phi) \Big] \bigg\} \\ \nonumber
 &\!\!=\!\!& 4 \bigg\{ R^2+ \rho_1^2+ (\rho_1^2+ \rho_2^2) \ln \rho_2 + 2 R^2 ( 1+ \ln R) \ln R \\ \nonumber && + \sum_{n=1}^\infty\Big[ \frac{ \rho_1^2}{n^2(n+1)} \Big( \frac{ \rho_1}{ \rho_2} \Big)^n - \frac{(\rho_1^2-\rho_2^2)}{n^2} \Big(\frac{\rho_1}{\rho_2}\Big)^n \Big] \\ \nonumber && + \sum_{n=2}^\infty \frac{1}{n^2(n-1)}\Big[\rho_2^2 \Big(\frac{\rho_1}{\rho_2}\Big)^n-R^2 \Big(\frac{\rho_1\rho_2}{R^2}\Big)^n\Big] \\ && + 2\rho_1 \rho_2 \ln (R/\rho_2) \cos \phi \bigg\} .
\end{eqnarray}
Making use of the series identities we have, 
\begin{widetext}
\begin{multline}
\label{eq: Edd}
E_{dd} = 4 {\rm Re} \Big[  R^2+ (\rho_1^2+ \rho_2^2) \ln \rho_2 + 2 R^2 \big( 1+ \ln R\big) \ln R -2 \rho_1 \rho_2 \cos \phi  \ln \Big( 1- \frac{ \rho_1}{ \rho_2} e^{i \phi} \Big)  + (\rho_1^2+\rho_2^2 ) \ln \Big( 1- \frac{ \rho_1}{ \rho_2} e^{i \phi} \Big) \\ + \rho_1 \rho_2 e^{i \phi}\ln \Big( 1- \frac{ \rho_1 \rho_2}{R^2} e^{i \phi} \Big) +R^2 {\rm Li}_2\Big(\frac{ \rho_1 \rho_2}{R^2} e^{i \phi} \Big)-   R^2\ln\ \Big( 1- \frac{ \rho_1 \rho_2}{R^2} e^{i \phi} \Big)+2\rho_1 \rho_2 \ln (R/\rho_2) e^{i \phi} \Big]
\end{multline}
\end{widetext}
Finally, we calculate the induced-direct terms,
\begin{eqnarray}
\nonumber
E^{12}_{id} &\!\! = \!\!&  \Big( \frac{ K^2_0 s_1 s_2 }{ (8 \pi)^2} \Big)^{-1}  \int_0^R dr~r~ \sigma_{kk}^{i1} (r) \sigma_{kk}^{d2} (r) \\ \nonumber 
 &\!\!=\!\!& -16 \bigg\{ \int_0^{\rho_2} dr ~r \Big[  \Big(\ln R+ \frac{1}{2} + \frac{\rho_1^2}{2 R^2} \Big) \big(\ln \rho_2 +1\big) \\ \nonumber && \ \ +  \frac{1}{2}\sum_{n=1}^\infty \frac{1}{n} \Big( \frac{ r^2 \rho_1}{ \rho_2 R^2 } \Big)^n \Big( \frac{n+1}{n} - \frac{\rho_1^2}{R^2} \Big)\cos (n \phi) \Big] \\ \nonumber && \int_{\rho_2}^R dr ~r \Big[  \Big(\ln R+ \frac{1}{2} + \frac{\rho_1^2}{2 R^2} \Big) \big(\ln r +1\big) \\ \nonumber && \ \ +  \frac{1}{2}\sum_{n=1}^\infty \frac{1}{n} \Big( \frac{  \rho_1 \rho_2}{  R^2 } \Big)^n \Big( \frac{n+1}{n} - \frac{\rho_1^2}{R^2} \Big)\cos (n \phi) \Big] \\ \nonumber &\!\!=\!\!& -4 \bigg\{\frac{1}{2R^2} (\rho_1^2+R^2+2R^2 \ln R)(\rho_2^2+R^2+2R^2 \ln R) \\ \nonumber && + \sum_{n=1}^\infty \Big(\frac{ \rho_1 \rho_2 }{R^2} \Big)^n  \Big[ \frac{ \rho_2^2}{n(n+1)}+ \frac{(R^2 - \rho_2^2)}{n} \Big] \\ && \ \ \times \Big( \frac{n+1}{n} - \frac{ \rho_1^2}{R^2} \Big) \cos( n\phi) \bigg\} .
 \end{eqnarray}
The corresponding term,  $E^{21}_{id}$, is obtained by interchanging $\rho_1$ and $\rho_2$ in $E^{12}_{id}$.  Combining both induced-direct terms,
\begin{eqnarray}
\nonumber
E_{id}  &\!\!=\!\!& E^{12}_{id}+E^{21}_{id} \\ \nonumber
&\!\!=\!\!& -4 \bigg\{\frac{1}{R^2} (\rho_1^2+R^2+2R^2 \ln R)(\rho_2^2+R^2+2R^2 \ln R) \\ \nonumber && + \sum_{n=1}^\infty \Big(\frac{ \rho_1 \rho_2 }{R^2} \Big)^n  \Big[ ( \rho_1^2+ \rho_2^2)\frac{(1-n)}{n^2} - 2 \Big(\frac{ \rho_1 \rho_2}{R^2} \Big)^2 \frac{R^2}{n(n+1)} \\ && + (2R^2-\rho_1^2-\rho_2^2)\frac{(n+1)}{n^2}  + 2\Big(\frac{ \rho_1 \rho_2}{R^2}\Big)^2 \frac{R^2}{n} \Big] \cos( n\phi) \bigg\} .
\end{eqnarray}
Evaluating the series, we have,
\begin{widetext}
\begin{multline}
\label{eq: Eid}
E_{id} = -4 \bigg\{\frac{1}{R^2} (\rho_1^2+R^2+2R^2 \ln R)(\rho_2^2+R^2+2R^2 \ln R) - 2 \frac{ \rho_2^2 \rho_2^2}{R^2} -2 \rho_1 \rho_2 e^{-i \phi} \ln \Big( 1- \frac{ \rho_1 \rho_2}{ R^2} e^{i \phi} \Big) \\  - 2\big(R^2 - \rho_1^2-\rho_2^2 \big) \ln \Big( 1- \frac{ \rho_1 \rho_2}{ R^2} e^{i \phi} \Big) + 2R^2 {\rm Li}_2 \Big(\frac{ \rho_1 \rho_2}{R^2} e^{i \phi} \Big) \bigg\}.
\end{multline}
\end{widetext}
Combining eq. (\ref{eq: Eii}), (\ref{eq: Edd}) and (\ref{eq: Eid}) we arrive at an expression for the elastic interaction between disclinations, $E_{int} = L K_0 s^2/(32 \pi) (E_{ii}+ E_{dd} + E_{id})$
\begin{multline}
\label{eq: Eint}
E_{int}(\rv_1, \rv_2)/L =\frac{K_0 s_1 s_2}{16 \pi}  \Big\{R^2\Big(1-\frac{\rho_1^2}{R^2} \Big)\Big(1- \frac{\rho_2^2}{R^2}\Big) \\ + |\rv_1-\rv_2|^2 \ln \Big[ \frac{|\rv_1-\rv_2|^2}{|\bar{\rv}_{12}-{\bf R}|^2} \Big] \Big\} ,
\end{multline} 
where 
\begin{eqnarray}
\nonumber
|\bar{\rv}_{12}-{\bf R}|^2 &\!\!=\!\!& R^2 +\frac{\rho_1^2 \rho_2^2}{R^2} - 2 \rho_1 \rho_2 \cos \phi \\
&\!\!=\!\!&  R^2\Big(1-\frac{\rho_1^2}{R^2} \Big)\Big(1- \frac{\rho_2^2}{R^2}\Big) +|\rv_1-\rv_2|^2  . 
\end{eqnarray}
Note that when either defect is pushed to the boundary, when $\rho_1 \to  R$ or $\rho_2 \to R$,  $|\bar{\rv}_{12}-{\bf R}|^2 = |\rv_1-\rv_2|^2$ and according to eq. (\ref{eq: Eint}), $E_{int}(\rv_1, \rv_2) =0$.  Also, note that in the limit that 2 defects of charge $s_1$ and $s_2$ fuse and $\rv_1 = \rv_2$, $E_{int}(\rv_1, \rv_1)/L = K_0 s_1 s_2 R^2 /(16 \pi) (1-\rho_1^2/R^2)^2$, which combined with the sum of the self-energies of the two elementary charges, yields the self energy of a disclination of total charge $s_1+s_2$.

\end{document}